\PassOptionsToPackage{unicode}{hyperref}
\PassOptionsToPackage{hyphens}{url}
\documentclass[
]{article}
\usepackage{amsmath,amssymb}
\usepackage{iftex}
\ifPDFTeX
  \usepackage[T1]{fontenc}
  \usepackage[utf8]{inputenc}
  \usepackage{textcomp} 
\else 
  \usepackage{unicode-math} 
  \defaultfontfeatures{Scale=MatchLowercase}
  \defaultfontfeatures[\rmfamily]{Ligatures=TeX,Scale=1}
\fi
\usepackage{lmodern}
\ifPDFTeX\else
\fi
\IfFileExists{upquote.sty}{\usepackage{upquote}}{}
\IfFileExists{microtype.sty}{
  \usepackage[]{microtype}
  \UseMicrotypeSet[protrusion]{basicmath} 
}{}
\makeatletter
\@ifundefined{KOMAClassName}{
  \IfFileExists{parskip.sty}{%
    \usepackage{parskip}
  }{
    \setlength{\parindent}{0pt}
    \setlength{\parskip}{6pt plus 2pt minus 1pt}}
}{
  \KOMAoptions{parskip=half}}
\makeatother
\usepackage{xcolor}
\usepackage[margin=1in]{geometry}
\usepackage{color}
\usepackage{fancyvrb}

\DefineVerbatimEnvironment{Highlighting}{Verbatim}{commandchars=\\\{\}}
\usepackage{framed}
\definecolor{shadecolor}{RGB}{248,248,248}
\newenvironment{Shaded}{\begin{snugshade}}{\end{snugshade}}

\newcommand{\AttributeTok}[1]{\textcolor[rgb]{0.13,0.29,0.53}{#1}}

\newcommand{\CommentTok}[1]{\textcolor[rgb]{0.56,0.35,0.01}{\textit{#1}}}

\newcommand{\ConstantTok}[1]{\textcolor[rgb]{0.56,0.35,0.01}{#1}}
\newcommand{\ControlFlowTok}[1]{\textcolor[rgb]{0.13,0.29,0.53}{\textbf{#1}}}

\newcommand{\DecValTok}[1]{\textcolor[rgb]{0.00,0.00,0.81}{#1}}
\newcommand{\DocumentationTok}[1]{\textcolor[rgb]{0.56,0.35,0.01}{\textbf{\textit{#1}}}}

\newcommand{\FloatTok}[1]{\textcolor[rgb]{0.00,0.00,0.81}{#1}}
\newcommand{\FunctionTok}[1]{\textcolor[rgb]{0.13,0.29,0.53}{\textbf{#1}}}

\newcommand{\NormalTok}[1]{#1}

\newcommand{\OtherTok}[1]{\textcolor[rgb]{0.56,0.35,0.01}{#1}}

\newcommand{\SpecialCharTok}[1]{\textcolor[rgb]{0.81,0.36,0.00}{\textbf{#1}}}

\newcommand{\StringTok}[1]{\textcolor[rgb]{0.31,0.60,0.02}{#1}}

\usepackage{longtable,booktabs,array}
\usepackage{calc} 
\usepackage{etoolbox}
\makeatletter
\patchcmd\longtable{\par}{\if@noskipsec\mbox{}\fi\par}{}{}
\makeatother
\IfFileExists{footnotehyper.sty}{\usepackage{footnotehyper}}{\usepackage{footnote}}
\makesavenoteenv{longtable}
\usepackage{graphicx}
\makeatletter
\def\maxwidth{\ifdim\Gin@nat@width>\linewidth\linewidth\else\Gin@nat@width\fi}
\def\maxheight{\ifdim\Gin@nat@height>\textheight\textheight\else\Gin@nat@height\fi}
\makeatother
\setkeys{Gin}{width=\maxwidth,height=\maxheight,keepaspectratio}
\makeatletter
\def\fps@figure{htbp}
\makeatother
\usepackage{fancyhdr}
\usepackage{booktabs}
\ifLuaTeX
  \usepackage{selnolig}  
\fi
\usepackage[authoryear]{natbib}
\IfFileExists{bookmark.sty}{\usepackage{bookmark}}{\usepackage{hyperref}}
\IfFileExists{xurl.sty}{\usepackage{xurl}}{} 
\hypersetup{
  pdftitle={A practical guide to causal discovery with cohort data},
  pdfauthor={Ryan M. Andrews; Ronja Foraita; Vanessa Didelez; Janine Witte; ;  Boston University School of Public Health;  Leibniz Institute for Prevention Research and Epidemiology - BIPS, Bremen;  University of Bremen},
  pdfkeywords={graphical models, software},
  hidelinks,
  pdfcreator={LaTeX via pandoc}}

\title{A practical guide to causal discovery with cohort data}
\author{Ryan M. Andrews\textsuperscript{1} \and Ronja Foraita\textsuperscript{2} \and Vanessa Didelez\textsuperscript{2,3} \and Janine Witte\textsuperscript{2,3} \and \hspace{10cm} \and \textsuperscript{1} Boston University School of Public Health \and \textsuperscript{2} Leibniz Institute for Prevention Research and Epidemiology - BIPS, Bremen \and \textsuperscript{3} University of Bremen}
\date{December 18, 2023}
\begin{document}
\maketitle
\begin{abstract}
In this guide, we present how to perform constraint-based causal discovery using three popular software packages: \texttt{pcalg} (with add-ons \texttt{tpc} and \texttt{micd}), \texttt{bnlearn}, and TETRAD. We focus on how these packages can be used with observational data and in the presence of mixed data (i.e., data where some variables are continuous, while others are categorical), a known time ordering between variables, and missing data. Throughout, we point out the relative strengths and limitations of each package, as well as give practical recommendations. We hope this guide helps anyone who is interested in performing constraint-based causal discovery on their data.
\end{abstract}

{
\setcounter{tocdepth}{2}
\tableofcontents
}
\hypertarget{introduction}{%
\section{Introduction}\label{introduction}}

Causal graphs provide an intuitive, yet theoretically founded, way of representing causal relationships between variables. The aim of causal discovery is to learn a graph from data. Broadly, causal discovery algorithms belong to one of three classes: (1) constraint-based approaches, e.g.~the PC algorithm \citep{causationpredictionsearch}, (2) score-based approaches, e.g.~greedy equivalence search \citep{chickering2002optimal}, or (3) approaches based on functional causal models, e.g.~LiNGAM \citep{Shimizuetal2006}. Constraint-based approaches, which are the focus of this guide, place the problem of causal discovery within the framework of a series of conditional independence hypothesis tests. These algorithms attempt to generate causal graphs with structures consistent with the conditional independence relationships found in the data. An advantage of constraint-based algorithms is that they are rather flexible in terms of the conditional independence tests that could be used. In addition, causal sufficiency can sometimes be relaxed \citep{causationpredictionsearch}. A notable disadvantage, however, of constraint-based algorithms is that any error(s) made are propagated through the algorithm (e.g., an edge mistakenly deleted early in the PC algorithm cannot be recovered later).

While the ``gold standard'' for discovering causal relationships is a randomized controlled trial, investigators often are interested in questions that can only be answered with observational data for practical and/or ethical reasons. Observational cohort data introduces a number of challenges for causal discovery. For example, missing data are ubiquitous in cohort studies because of dropout, incomplete assessments, or missed visits. When data are not missing completely at random, conditional independences found in a complete case analysis (i.e., after deleting any incomplete data rows) may differ from those that would be found if the data were complete, even if the sample size is large. Therefore, causal discovery algorithms may generate incorrect graphs in the presence of missing data. When the cohort is followed over time, this can introduce additional complexities. For example, it is known that events in the future cannot cause those in the past, and this so-called temporal ordering should be accounted for in the causal discovery algorithms. Further, cohort studies usually collect information on many different variables that can be continuous, binary, ordinal, or categorical; however, many causal discovery algorithms cannot accommodate such ``mixed'' data and require that all variables considered either be continuous or binary.

While there are many introductory papers on causal discovery \citep{KalischBuehlmann2014, SpirtesZhang2016, HeinzeDemletal2018, malinsky2018causal, glymour2019review}, none focuses on the actual implementation of the various algorithms in standard software. Therefore, information on the available causal discovery software packages are scattered across the literature (usually as software manual or \texttt{R} package vignette), which makes it more difficult for users to compare and contrast them. In this paper, we aim to provide a practical guide for carrying out causal discovery via three of the most popular software implementations: the Java-based TETRAD software package \citep{TETRAD}, the \texttt{R} package \texttt{bnlearn} \citep{R_bnlearn}, and the \texttt{R} package \texttt{pcalg} \citep{R_pcalg1, R_pcalg2} with its add-ons \texttt{tpc} \citep{Witte2021} and \texttt{micd} \citep{Foraitaetal2020, Witteetal2021}. For simplicity, we will focus only on the constraint-based PC algorithm and how it is implemented in each software package.

The paper is organized as follows. In Section 2, we provide a brief introduction to basic causal discovery terminology, followed by introductions to \texttt{pcalg}, \texttt{bnlearn}, and TETRAD. Here, we focus on the basic function(s) within each package that are used to implement the PC algorithm and their required inputs. In Section 3.1, we provide background on the simulated data we are using in the examples throughout the guide. In Sections 3.2, 3.3, and 3.4, we expand on Section 2 by illustrating how each causal discovery software package can be used in the presence of mixed data, time ordering, and missing data, respectively. A special focus is on how the new packages \texttt{tpc} and \texttt{micd} expand the functionality of \texttt{pcalg}. The paper concludes with a brief discussion of causal effect estimation after learning a causal graph in Section 3.5, and some general advice and recommendations in Section 4.

\hypertarget{causal-discovery-terminology-and-software}{%
\section{Causal discovery terminology and software}\label{causal-discovery-terminology-and-software}}

\hypertarget{terminology}{%
\subsection{Terminology}\label{terminology}}

At a basic level, the goal of causal discovery is to generate potential causal graphs from data, statistical algorithms, and background knowledge. A graph consists of \textit{nodes} and \textit{edges}, which correspond to the variables in the graph and the arrows between these variables, respectively. A graph that consists only of directed edges and contains no cycles (i.e., one cannot start at a node and arrive back at the same node on the graph by following the direction of the edges on the graph) is called a \textit{directed acyclic graph} (DAG). A causal DAG is a DAG in which the presence of an edge between nodes indicates a belief that there is a direct causal relation; the absence of an edge between nodes indicates the absence of a direct causal relation. Another feature of a causal DAG is that all causes (even if unmeasured) are on the graph. For causal discovery, we therefore need to decide whether we believe that all relevant variables have been measured, in the sense that if two variables in the data share a common cause, then that common cause needs to be included in the data as well. This assumption is called \textit{causal sufficiency} and is required by PC and many other causal discovery algorithms. If causal sufficiency cannot be assumed, other constraint-based algorithms such as the `Fast Causal Inference' (FCI) algorithm \citep{causationpredictionsearch} can still be applied, but they produce a less informative output.

Constraint-based causal discovery algorithms such as PC, on which we focus in this guide, build the graph based on conditional independences in the data, which are inferred using conditional independence testing. Since several DAGs can imply the same set of conditional independences (they are then said to be \textit{Markov equivalent} and form a \textit{Markov equivalence class}), PC is not able to distinguish between them. Therefore, even under ideal circumstances (infinite sample size, all assumptions are true), the output of PC is not a DAG, but a \textit{completed partially directed acyclic graph} (CPDAG) containing both directed and undirected edges. An undirected edge indicates that the algorithm determined that two nodes were related to each other, but it could not determine which node causes the other. If background knowledge is used during the search (e.g.~in the form of required edges or a partial node ordering), then the output of PC is an estimated \textit{maximally oriented partially directed acyclic graph} (MPDAG), which has a slightly different interpretation from an estimated CPDAG.

Depending on the software used, the output of PC may also contain bi-directed edges (in which case, the output is no longer a CPDAG or MPDAG). These occur when different conditional independence tests produce conflicting results. In contrast to the undirected edges, whose orientation cannot be determined even with an infinite sample size, the bi-directed edges always point to conflicts that are due to the finite data. Below, we discuss what options for conflict resolution are offered by the different software packages.

An important parameter of PC and similar algorithms is the p-value threshold, usually called \texttt{alpha}. Each conditional independence test performed by the algorithm returns a p-value, which is then used in order to make a yes-no decision: If the p-value of a particular test is smaller than \texttt{alpha}, then this is treated as a dependence, otherwise an independence is concluded. Since concluding independence usually means deleting an edge from the estimated graph, \texttt{alpha} controls its sparsity, where a smaller value leads to a sparser output.

\hypertarget{r-package-pcalg-and-add-ons}{%
\subsection{\texorpdfstring{R-package \texttt{pcalg} and add-ons}{R-package pcalg and add-ons}}\label{r-package-pcalg-and-add-ons}}

First released in 2009, the \texttt{pcalg} package \citep{R_pcalg1, R_pcalg2} contains many functions to implement causal discovery algorithms and estimate causal effects. In this guide, we consider version 2.7-3. (see www.cran.r-project.org/web/packages/pcalg/index.html for the latest version, along with vignettes). In \texttt{pcalg}, the PC algorithm is implemented in the \texttt{pc()} function, which takes the following inputs:

\(~\)

\begin{Shaded}
\begin{Highlighting}[]
\FunctionTok{pc}\NormalTok{(suffStat, indepTest, alpha, labels, p,}
   \AttributeTok{fixedGaps =} \ConstantTok{NULL}\NormalTok{, }\AttributeTok{fixedEdges =} \ConstantTok{NULL}\NormalTok{, }\AttributeTok{NAdelete =} \ConstantTok{TRUE}\NormalTok{, }\AttributeTok{m.max =} \ConstantTok{Inf}\NormalTok{,}
   \AttributeTok{u2pd =} \FunctionTok{c}\NormalTok{(}\StringTok{"relaxed"}\NormalTok{, }\StringTok{"rand"}\NormalTok{, }\StringTok{"retry"}\NormalTok{),}
   \AttributeTok{skel.method =} \FunctionTok{c}\NormalTok{(}\StringTok{"stable"}\NormalTok{, }\StringTok{"original"}\NormalTok{, }\StringTok{"stable.fast"}\NormalTok{),}
   \AttributeTok{conservative =} \ConstantTok{FALSE}\NormalTok{, }\AttributeTok{maj.rule =} \ConstantTok{FALSE}\NormalTok{, }\AttributeTok{solve.confl =} \ConstantTok{FALSE}\NormalTok{, }
   \AttributeTok{numCores =} \DecValTok{1}\NormalTok{, }\AttributeTok{verbose =} \ConstantTok{FALSE}\NormalTok{)}
\end{Highlighting}
\end{Shaded}

\(~\)

At a high level, the user must provide \texttt{pc()} with a function performing the conditional independence tests (\texttt{indepTest}) along with a p-value threshold \texttt{alpha} and a summary of the dataset (\texttt{suffStat}, `sufficient statistic'). Depending on the chosen \texttt{indepTest} function, \texttt{suffStat} is a list of summary statistics or just the dataset itself. Users may also write their own \texttt{indepTest} function or use one of the functions from the \texttt{micd} package for handling missing data. The variable names are provided in \texttt{labels}; the parameter \texttt{p} can then be ignored.

Forbidden and required adjacencies may be specified in \texttt{fixedGaps} and \texttt{fixedEdges}, respectively. If two nodes \(A\) and \(B\) are forbidden to be adjacent, then no edge of any orientation will appear between them in the estimated CPDAG or MPDAG, but it is not possible to specify that e.g.~\(A\rightarrow B\) is forbidden but \(A\leftarrow B\) is allowed. If an adjacency between nodes \(A\) and \(B\) is required, an edge between them is forced into the estimated CPDAG or MPDAG, but the orientation is determined by the algorithm.

The \texttt{NAdelete} input determines what happens if the \texttt{indepTest} function returns \texttt{NA}. If \texttt{NAdelete\ =\ TRUE}, which is the default option, then each \texttt{NA} is treated as if the test returned a p-value of 1, which usually means that an edge is deleted.

The next inputs determine exactly which version of PC is used. Setting \texttt{m.max} to a small integer value has the effect that only conditional independence tests up to a certain order (i.e., number of variables in the conditioning set) are performed. This can be useful if the runtime would be prohibitively long otherwise, or if the higher-order tests are believed to be too unreliable. We recommend sticking to the default \texttt{m.max\ =\ Inf} (i.e., no restrictions) at least to start with. We further recommend setting \texttt{u2pd\ =\ "relaxed",\ skel.method\ =\ "stable",\ maj.rule\ =\ TRUE,\ solve.confl\ =\ TRUE} in order to obtain an output that does not depend on the order in which the variables appear in the dataset (see \citet{ColomboMaathuis2014} for details on the problem and the solution). This has the effect that edges remain undirected whenever there is conflicting information about their orientation. When plotting the estimated graph using the \texttt{pcalg::plot()} function, these edges are shown as bi-directed edges; we recommend checking the adjacency matrix to verify which edges are actually undirected and which are bi-directed. By specifying \texttt{skel.method\ =\ "stable.fast"} and setting \texttt{numCores} to an integer value larger than 1, parts of \texttt{pc()} can be run in parallel on a multi-core machine, thus saving runtime without altering the output. Finally, \texttt{verbose\ =\ TRUE} prints detailed log information into the console. Additional details on the \texttt{pc()} function can be found by typing \texttt{help(pc)} into the \texttt{R} console.

The \texttt{tpc()} function from the \texttt{tpc} package, available from CRAN and www.github.com/bips-hb/tpc, is a modified version of \texttt{pc()} with the following input structure:

\(~\)

\begin{Shaded}
\begin{Highlighting}[]
\FunctionTok{tpc}\NormalTok{(suffStat, indepTest, alpha, labels, p,}
    \AttributeTok{forbEdges =} \ConstantTok{NULL}\NormalTok{, }\AttributeTok{m.max =} \ConstantTok{Inf}\NormalTok{,}
    \AttributeTok{conservative =} \ConstantTok{FALSE}\NormalTok{, }\AttributeTok{maj.rule =} \ConstantTok{TRUE}\NormalTok{,}
    \AttributeTok{tiers =} \ConstantTok{NULL}\NormalTok{, }\AttributeTok{context.all =} \ConstantTok{NULL}\NormalTok{, }\AttributeTok{context.tier =} \ConstantTok{NULL}\NormalTok{,}
    \AttributeTok{verbose =} \ConstantTok{FALSE}\NormalTok{)}
\end{Highlighting}
\end{Shaded}

\(~\)

In contrast to \texttt{pc()}, \texttt{tpc()} does not support parallel computing, and \texttt{u2pd\ =\ "relaxed",\ skel.method\ =\ "stable",\ solve.confl\ =\ TRUE} is chosen by default without alternative options. In \texttt{forbEdges}, users can specify directed edges that should not appear in the estimated CPDAG or MPDAG, e.g., forbid \(A\rightarrow B\) (but allow \(A\leftarrow B\)), or forbid both, with the effect that there will be no edge between \(A\) and \(B\) in the estimated graph. This is analogous to the \texttt{blacklist} option of \texttt{bnlearn} discussed in the next section.

The most important difference to \texttt{pc()} is the availability of the \texttt{tiers}, \texttt{context.all}, and \texttt{context.tier} options, which are explained below in Section 3.3.1.

\hypertarget{r-package-bnlearn}{%
\subsection{\texorpdfstring{R-package \texttt{bnlearn}}{R-package bnlearn}}\label{r-package-bnlearn}}

The \texttt{bnlearn} (Bayesian network learning and inference) package \citep{R_bnlearn} was first released in 2007. In this guide, we are using version 4.6.1. See the homepage (www.bnlearn.com) for helpful information. The PC algorithm is implemented through the \texttt{pc.stable()} function:

\begin{Shaded}
\begin{Highlighting}[]
\FunctionTok{pc.stable}\NormalTok{(x, }\AttributeTok{cluster =} \ConstantTok{NULL}\NormalTok{, }\AttributeTok{whitelist =} \ConstantTok{NULL}\NormalTok{, }\AttributeTok{blacklist =} \ConstantTok{NULL}\NormalTok{, }\AttributeTok{test =} \ConstantTok{NULL}\NormalTok{,}
          \AttributeTok{alpha =} \FloatTok{0.05}\NormalTok{, }\AttributeTok{B =} \ConstantTok{NULL}\NormalTok{, }\AttributeTok{max.sx =} \ConstantTok{NULL}\NormalTok{, }\AttributeTok{debug =} \ConstantTok{FALSE}\NormalTok{,}
          \AttributeTok{undirected =} \ConstantTok{FALSE}\NormalTok{)}
\end{Highlighting}
\end{Shaded}

The inputs to \texttt{pc.stable()} have some similarities and many differences compared to the \texttt{pc()} function in \texttt{pcalg}. For example, the \texttt{pc.stable()} function also allows the user to set a p-value threshold (\texttt{alpha}), a maximal conditioning set size (\texttt{max.sx}), an option for parallel computing (\texttt{cluster}), and an option for detailed log output (\texttt{debug\ =\ TRUE}). On the other hand, \texttt{pc.stable()} allows the user to input the data directly into the function (\texttt{x}) rather than the data's sufficient statistics. The conditional independence test (\texttt{test}) to be used can be chosen from among several inbuilt options; it is not possible to provide a user-built function.

The \texttt{whitelist} and \texttt{blacklist} inputs allow users to specify sets of edges that should and should not be included in the estimated CPDAG or MPDAG, respectively. In contrast to \texttt{pc()} from the \texttt{pcalg} package, \texttt{pc.stable()} also considers the specified orientations, making it possible e.g.~to forbid \(A\rightarrow B\) but not \(A\leftarrow B\), or to specifically force \(A\leftarrow B\) into the output.

Overall, \texttt{pc.stable()} offers fewer options for fine-tuning the algorithm compared to \texttt{pc()} from the \texttt{pcalg} package or TETRAD (see next section). Conflicts are dealt with by comparing p-values, where decisions corresponding to smaller p-values are given preference over those with larger p-values. Using this heuristic, \texttt{pc.stable()} always produces an order-independent output. However, there is no theoretical reason why the orientations obtained this way should be more likely than the alternative orientations not chosen.

Additional details on the \texttt{pc.stable()} function can be found by typing \texttt{help(pc.stable)} into the \texttt{R} console.

\hypertarget{tetrad}{%
\subsection{TETRAD}\label{tetrad}}

TETRAD is a Java program for causal discovery and structural equation modeling developed by researchers from the Center for Causal Discovery (CCD) \citep{TETRAD, Cooperetal2015} (www.ccd.pitt.edu/tools). Unlike \texttt{pcalg} and \texttt{bnlearn}, TETRAD is designed for users who prefer a graphical user interface (GUI), and it is possible to perform causal discovery without writing any code, but R and Python interfaces are available from the CCD website as well. In this guide, we used GUI-based TETRAD version 6.9.0.

\begin{figure}
\centering
\includegraphics{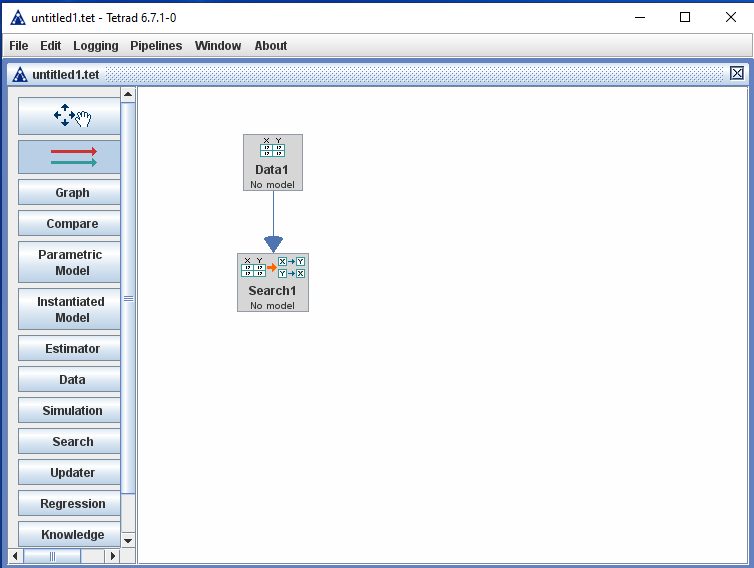}
\caption{\label{fig:tetrad-load-data}Loading data into TETRAD.}
\end{figure}

To implement the PC algorithm in TETRAD, one must first add a ``Data'' and ``Search'' box to the GUI and connect them with an arrow, see Figure \ref{fig:tetrad-load-data}.

\begin{figure}
\centering
\includegraphics{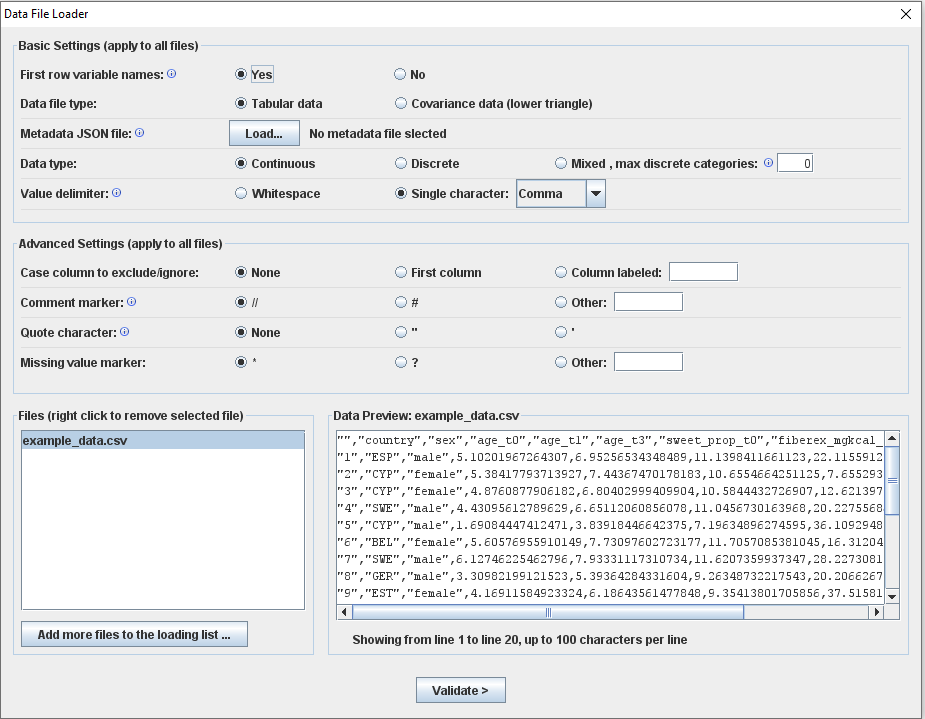}
\caption{\label{fig:tetrad-data-features}Specifying various data features, including missing data markers.}
\end{figure}

Then, the user can double click on the ``Data'' icon, and select \emph{File} \(\rightarrow\) \emph{Load} to load the data into TETRAD. This brings up another GUI that allows for the specification of various data features, like missing data markers, as shown in Figure \ref{fig:tetrad-data-features}.

The user must validate the data with TETRAD before it is loaded. During the validation process, TETRAD checks for potential issues, like accidentally including a bookkeeping row (e.g., the variable names) as a data observation or not correctly labeling missing data values. If issues are found, TETRAD will display the row number(s) and issue(s) found so they can be corrected. After all validation checks are passed, the data can be loaded and search can be performed.

\begin{figure}
\centering
\includegraphics{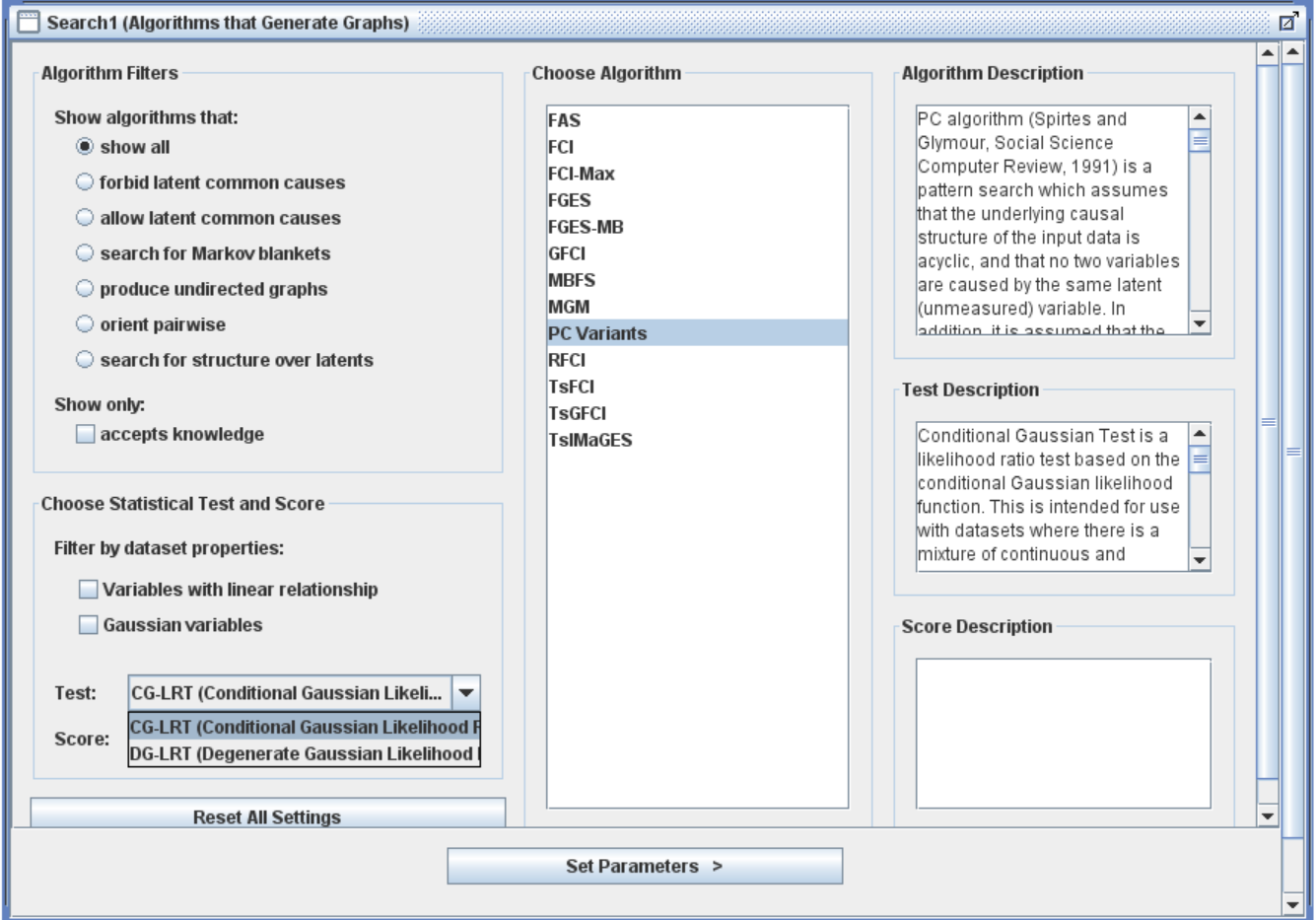}
\caption{\label{fig:tetrad-search-algs}Selecting a search algorithm in TETRAD.}
\end{figure}

\begin{figure}
\centering
\includegraphics{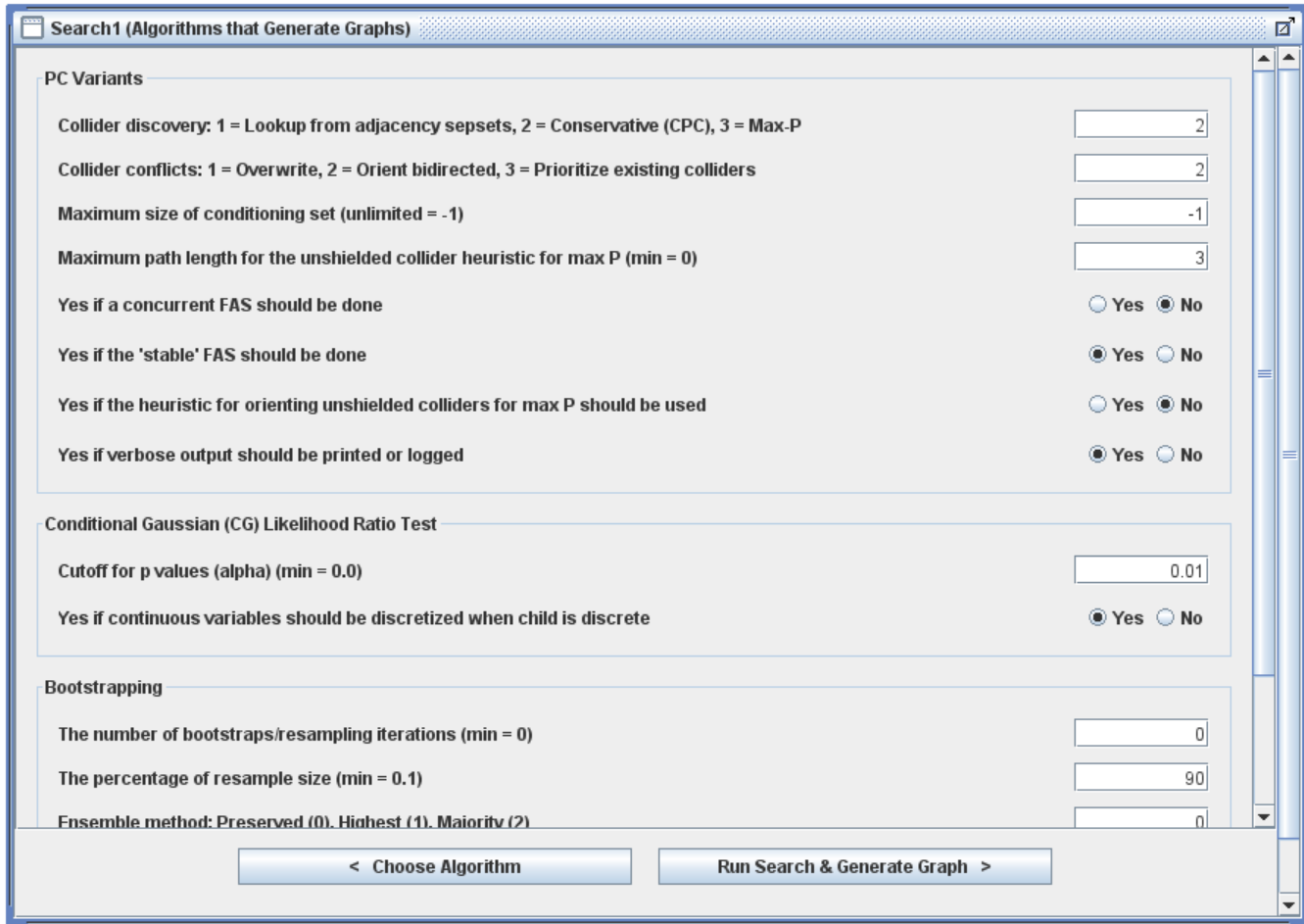}
\caption{\label{fig:tetrad-search-options}Specifying various options related to the causal search.}
\end{figure}

When setting up a causal search, the first thing that must be done after clicking on the ``Search'' icon is selecting the search algorithm, see Figure \ref{fig:tetrad-search-algs}. Here, we tell TETRAD that we want to use the PC algorithm. In addition, we specify the conditional independence test to be used. The options to choose from depend on the measurement scales of the data; in Figure \ref{fig:tetrad-search-algs}, the options for a dataset containing both numeric and discrete variables are shown. After choosing the algorithm, users can further specify options related to how the search should be performed, including how conflicts between edges are handled, maximal conditioning sets, p-value thresholds, and bootstrapping options, see Figure \ref{fig:tetrad-search-options}.

More information on all options can be obtained by hovering the mouse over text in the ``Search'' box or by consulting the manual at www.github.com/cmu-phil/tetrad.

\hypertarget{comparison}{%
\subsection{Comparison}\label{comparison}}

Table \ref{tab:software} provides an overview of the capabilities of \texttt{pcalg}, \texttt{bnlearn}, and TETRAD with respect to common issues with cohort data. Details are discussed below after we introduced our synthetic cohort data.

\begin{table}
    \caption{Overview of \texttt{pcalg}, \texttt{tpc}, \texttt{micd}, \texttt{bnlearn}, and TETRAD with respect to their ability to handle mixed measurement scales, time-ordered variables, and missing data.}
    \label{tab:software}
    \begin{tabular}{llll}
        \toprule
        package & mixed data & time-ordering & missing data \\
        \midrule
    \texttt{pcalg}/\texttt{tpc}/\texttt{micd} & \multicolumn{1}{p{4cm}}{Conditional Gaussian test (\texttt{mixCItest} from package \texttt{micd})} & \multicolumn{1}{p{4cm}}{\texttt{tiers} argument in \texttt{tpc}} & \multicolumn{1}{p{4cm}}{multiple imputation and test-wise deletion (using test functions from the \texttt{micd} package)} \\
    \midrule
    \texttt{bnlearn} & \multicolumn{1}{p{4cm}}{Conditional Gaussian test (\texttt{test = "mi-cg"})} & \multicolumn{1}{p{4cm}}{\texttt{blacklist} argument} & \multicolumn{1}{p{4cm}}{test-wise deletion is automatically applied}\\
    \midrule
    TETRAD & \multicolumn{1}{p{4cm}}{Conditional Gaussian and Degenerate Gaussian test} & \multicolumn{1}{p{4cm}}{via "tier" specification} & \multicolumn{1}{p{4cm}}{test-wise deletion is automatically applied} \\
        \bottomrule
    \end{tabular}
\end{table}

\hypertarget{synthetic-cohort-data-example}{%
\section{Synthetic cohort data example}\label{synthetic-cohort-data-example}}

\hypertarget{ideficsi.family-cohort}{%
\subsection{IDEFICS/I.Family cohort}\label{ideficsi.family-cohort}}

To make our examples easily replicable, we constructed synthetic data which is included in the \texttt{tpc} package and which is modeled after real cohort data. The Identification and Prevention of Dietary and Lifestyle-induced Health Effects in Children and Infants (IDEFICS) study was a European prospective cohort study designed to investigate the relationship between non-communicable chronic diseases and dietary, lifestyle, behavioral, and socioeconomic factors among preschool and primary school aged children \citep{ahrens2006understanding}. The I.Family study \citep{ahrens2017cohort} began as an extension of the IDEFICS study, with its baseline being the second IDEFICS follow-up visit. The cohort includes longitudinal data across the key development phases from toddler to adult.

\begin{figure}
\centering
\includegraphics{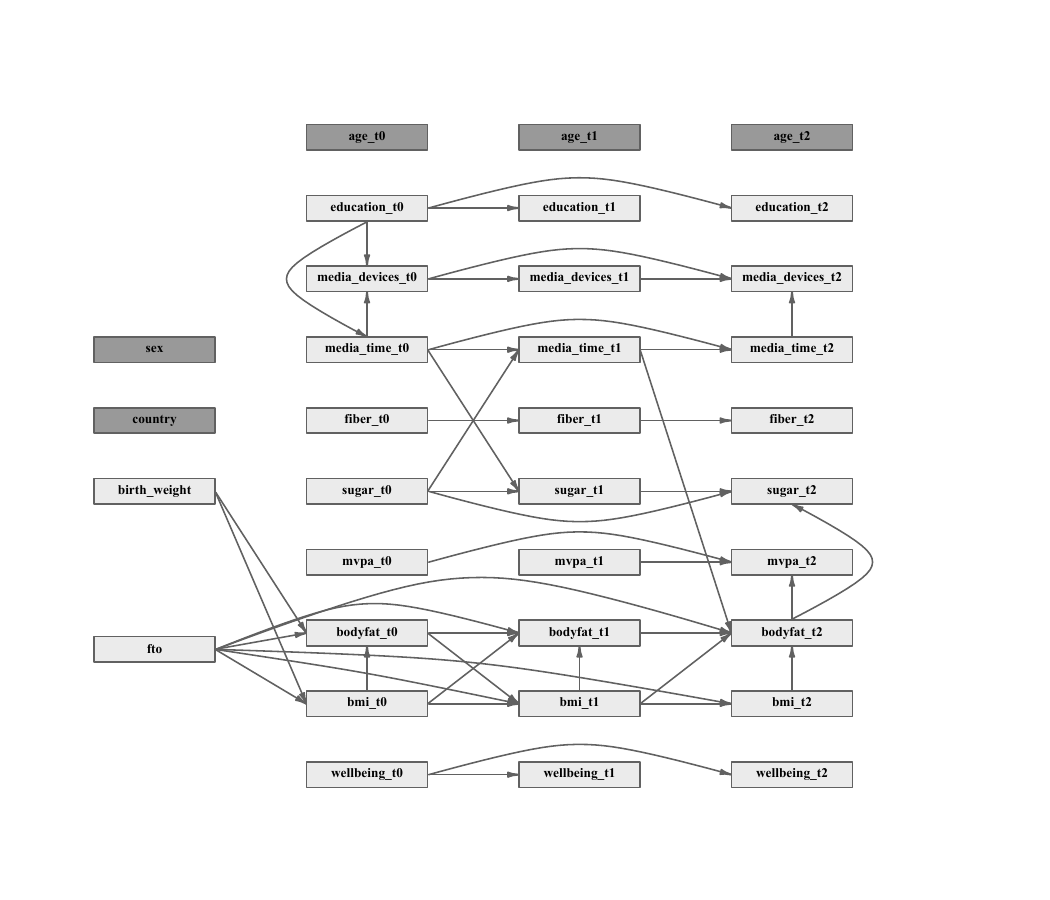}
\caption{\label{fig:causal-dag}Assumed causal mechanisms and associations in the simulated data.}
\end{figure}

For this tutorial, we have mimicked IDEFICS/I.Family data in the following way:
From the list of all IDEFICS/I.Family variables, we chose two baseline characteristics (sex and country), one genetic factor (high-risk alleles of the fat mass and obesity associated (FTO) gene), one early life factor (birth weight) and 10 variables that were measured in each of the three waves. See Table \ref{tab:vars} for more information on all chosen variables. We specified the causal DAG in Figure \ref{fig:causal-dag} based on assumed causal mechanisms and on associations in the data. For the rest of this paper, we will refer to the DAG in Figure
\ref{fig:causal-dag} as the \textit{true} DAG, as it is the model generating the data used for illustration.

We equipped the true DAG with a joint distribution by specifying how each variable \(X\) is generated given the values of its parents. For multinomial \(X\), we assumed a multinomial distribution; for continuous \(X\), a main effects linear regression; for ordered categorical \(X\), a main effects ordered logistic model; for count data \(X\), a main effects Poisson model. In order to have realistic coefficients, we estimated them from the IDEFICS/I.Family data. Using this data-generating process, we generated 5000 artificial observations (dataset \texttt{dat\_cohort} in the \texttt{tpc} package). We compared the marginal distributions of all variables to those in the real data and found that the means and standard deviations were similar. The \texttt{tpc} package also contains a dataset \texttt{dat\_cohort\_dis} where the continuous data have been discretized, and a datasets \texttt{dat\_cohort\_mis} (see Section 3.4). All three datasets will be used as examples in this guide. A causal discovery analysis using the real IDEFICS data is performed in \citet{Foraitaetal2021}.

\begin{table}
    \caption{Variables in example dataset. Transformations were chosen such that the marginal distribution after transformation was roughly symmetric.}
    \label{tab:vars}
    \begin{tabular}{lll}
        \toprule
        name & notes & scale \\
        \midrule
    \multicolumn{3}{l}{\textbf{baseline characteristics}} \\
    country & 1=ITA, 2=EST, 3=CYP, 4=BEL, 5=SWE, 6=GER, & categorical (nominal)\\
     & 7=HUN, 8=ESP &\bigskip\\
    sex & 1=male, 2=female & binary\\
    \multicolumn{3}{l}{\textbf{genetic factors}} \\
    fto & number of copies of a high-risk allele of the FTO gene & ordinal/count\bigskip\\
    \multicolumn{3}{l}{\textbf{early life factors}} \\
    birth\_weight & birth weight in g &  continuous\bigskip\\
    \multicolumn{3}{l}{\textbf{time-dependent variables}} \\
    age & age in years & continuous\\
    bmi & body mass index (BMI) z-score adjusted for sex and age & continuous\\
    bodyfat & \% bodyfat & continuous\\
    education & parent's highest education, 1=low, 2=medium, 3=high & ordinal\\
    fiber & fiber intake in log(mg/kcal) & continuous \\
    media\_devices & number of audiovisual media in the child's bedroom & count\\
    media\_time & use of audiovisual media in log(hours/week) & continuous \\
    mvpa & moderate to vigorous physical activity & continuous in sqrt(min/day) \\
    sugar & sugar intake score & continuous \\
    wellbeing & well-being score & continuous \\
        \bottomrule
    \end{tabular}
\end{table}

\hypertarget{mixed-measurement-scales}{%
\subsection{Mixed measurement scales}\label{mixed-measurement-scales}}

Cohort studies collect a mix of continuous or discrete measurements. For example, a typical epidemiologic cohort study will record participants' age (continuous), race (categorical), and sex (binary), in addition to many other variables.

Using ``mixed data,'' where variables are not all on the same measurement scale, in causal discovery, requires suitable tests for the mixed scales. Independence between purely continuous variables is usually tested with a partial correlation test, e.g.,~Fisher's z-test or the partial correlation t-test, while independence testing between discrete variables is carried out with a \(G^2\) or \(X^2\) test. There is no general consensus on which test to use for mixed data. \citet{AndrewsRamseyCooper2018} proposed a likelihood ratio test (`Conditional Gaussian test') based on the Conditional Gaussian assumption, and \citet{AndrewsRamseyCooper2019} suggested to transform the categorical variables into 0-1 dummy variables and treat them as normally distributed afterwards (`Degenerate Gaussian test'). Other, non-parametric options exist but are more involved and not covered here.

To illustrate how each causal discovery implementation handles mixed data, it suffices to consider cross-section (i.e., baseline) data. Later, longitudinal data will be considered.

\begin{Shaded}
\begin{Highlighting}[]
\DocumentationTok{\#\# install required packages}
\CommentTok{\#install.packages("pcalg")}
\CommentTok{\#install.packages("bnlearn")}
\CommentTok{\#install.packages("tpc")}

\DocumentationTok{\#\# load required packages}
\FunctionTok{library}\NormalTok{(bnlearn)}
\FunctionTok{library}\NormalTok{(pcalg)}
\FunctionTok{library}\NormalTok{(micd)}
\FunctionTok{library}\NormalTok{(tpc)}

\DocumentationTok{\#\# load cohort data and create cross{-}sectional version of the dataset}
\FunctionTok{data}\NormalTok{(}\StringTok{"dat\_cohort"}\NormalTok{)}
\NormalTok{dat\_cross }\OtherTok{\textless{}{-}}\NormalTok{ dat\_cohort[ ,}\DecValTok{1}\SpecialCharTok{:}\DecValTok{14}\NormalTok{]}

\DocumentationTok{\#\# load discretized data and create cross{-}sectional version of the dataset}
\FunctionTok{data}\NormalTok{(}\StringTok{"dat\_cohort\_dis"}\NormalTok{)}
\NormalTok{dat\_cross\_dis }\OtherTok{\textless{}{-}}\NormalTok{ dat\_cohort\_dis[ ,}\DecValTok{1}\SpecialCharTok{:}\DecValTok{14}\NormalTok{]}
\end{Highlighting}
\end{Shaded}

\hypertarget{pcalg-and-add-ons}{%
\subsubsection{\texorpdfstring{\texttt{pcalg} and add-ons}{pcalg and add-ons}}\label{pcalg-and-add-ons}}

\begin{figure}
\centering
\includegraphics{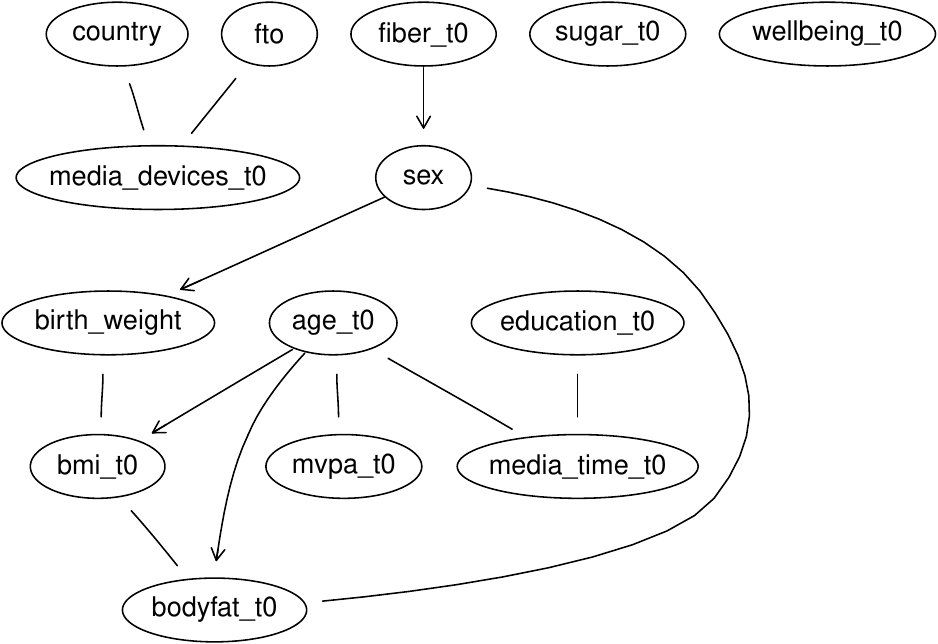}
\caption{\label{fig:pc-alg-mixed-cross}Output of the PC algorithm on our synthetic mixed data, as run by the \texttt{pc()} function in \texttt{pcalg}.}
\end{figure}

As mentioned previously, a required input to the \texttt{pc()} function is \texttt{indepTest}, which specifies how conditional independence tests will be carried out. A conditional independence test for mixed data, \texttt{mixCItest}, is included in the \texttt{pcalg} add-on \texttt{micd} \citep{Foraitaetal2020, Witteetal2021}, available at www.github.com/bips-hb/micd and via CRAN. It implements the likelihood-ratio test proposed by \citet{AndrewsRamseyCooper2018}, which assumes that the variables in the test follow a Conditional Gaussian distribution. The following code runs \texttt{pc} on our synthetic mixed data; the resulting graph estimate is shown in Figure \ref{fig:pc-alg-mixed-cross}.

\begin{Shaded}
\begin{Highlighting}[]
\NormalTok{pcalg\_fit\_mix }\OtherTok{\textless{}{-}} \FunctionTok{pc}\NormalTok{(}\AttributeTok{suffStat =}\NormalTok{ dat\_cross, }\AttributeTok{indepTest =}\NormalTok{ mixCItest, }\AttributeTok{alpha =} \FloatTok{0.01}\NormalTok{,}
                    \AttributeTok{labels =} \FunctionTok{colnames}\NormalTok{(dat\_cross), }\AttributeTok{u2pd=}\StringTok{"relaxed"}\NormalTok{,}
                    \AttributeTok{skel.method =} \StringTok{"stable"}\NormalTok{, }\AttributeTok{maj.rule =} \ConstantTok{TRUE}\NormalTok{, }\AttributeTok{solve.confl =} \ConstantTok{TRUE}\NormalTok{)}
\NormalTok{mygraph }\OtherTok{\textless{}{-}} \ControlFlowTok{function}\NormalTok{(pcgraph)\{}
\NormalTok{  g }\OtherTok{\textless{}{-}} \FunctionTok{as.bn}\NormalTok{(pcgraph, }\AttributeTok{check.cycles =} \ConstantTok{FALSE}\NormalTok{)}
  \FunctionTok{graphviz.plot}\NormalTok{(g, }\AttributeTok{shape =} \StringTok{"ellipse"}\NormalTok{)}
\NormalTok{\}}
\FunctionTok{mygraph}\NormalTok{(pcalg\_fit\_mix)}
\end{Highlighting}
\end{Shaded}

Alternatively, one could bin all the continuous variables into discrete groups, and then use the \texttt{disCItest()} function from the \texttt{pcalg} package, which implements a \(G^2\) test \citep{R_pcalg1}. In the following code, we use a version of our simulated dataset in which all variables have been discretized:

\(~\)

\begin{Shaded}
\begin{Highlighting}[]
\NormalTok{suffStat }\OtherTok{\textless{}{-}} \FunctionTok{list}\NormalTok{(}\AttributeTok{dm =}\NormalTok{ dat\_cross\_dis, }\AttributeTok{adaptDF =} \ConstantTok{FALSE}\NormalTok{)}
\NormalTok{pcalg\_fit\_dis }\OtherTok{\textless{}{-}} \FunctionTok{pc}\NormalTok{(}\AttributeTok{suffStat =}\NormalTok{ suffStat, }\AttributeTok{indepTest =}\NormalTok{ disCItest, }\AttributeTok{alpha =} \FloatTok{0.01}\NormalTok{,}
                    \AttributeTok{labels =} \FunctionTok{colnames}\NormalTok{(dat\_cross\_dis), }\AttributeTok{u2pd =} \StringTok{"relaxed"}\NormalTok{,}
                    \AttributeTok{skel.method =} \StringTok{"stable"}\NormalTok{, }\AttributeTok{maj.rule =} \ConstantTok{TRUE}\NormalTok{, }\AttributeTok{solve.confl =} \ConstantTok{TRUE}\NormalTok{)}
\FunctionTok{mygraph}\NormalTok{(pcalg\_fit\_dis)}
\end{Highlighting}
\end{Shaded}

\begin{figure}
\centering
\includegraphics{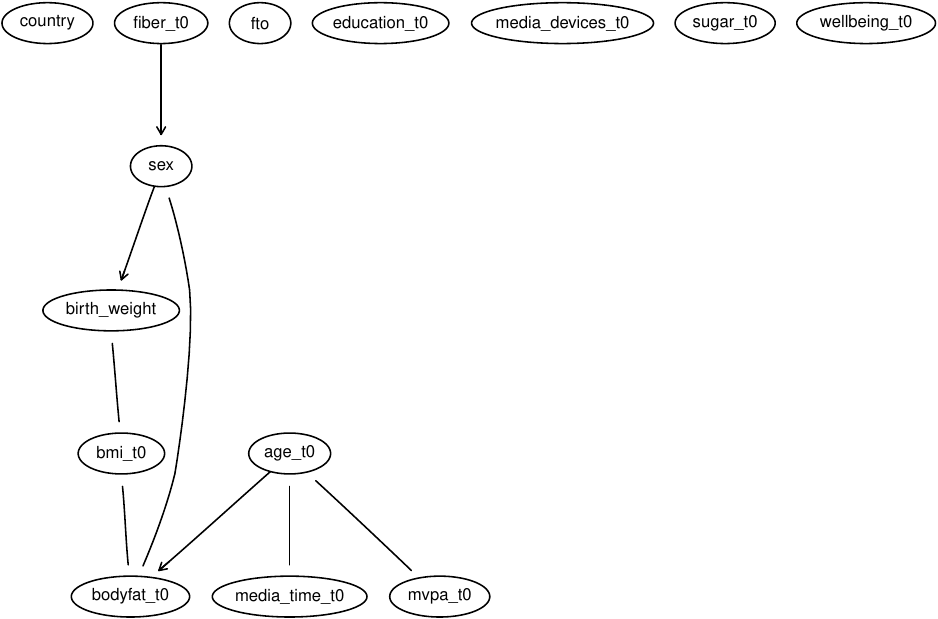}
\caption{\label{fig:pc-alg-mixed-discretized}Output from the \texttt{pc()} function in \texttt{pcalg}, but fit to data where all variables are discrete.}
\end{figure}

\(~\)

Figure \ref{fig:pc-alg-mixed-discretized} shows the result. Here, we see that considerable fewer edges were detected compared to Figure \ref{fig:pc-alg-mixed-cross}, presumably due to the lower power of the test that does not presume a linear dependence.

\hypertarget{bnlearn}{%
\subsubsection{\texorpdfstring{\texttt{bnlearn}}{bnlearn}}\label{bnlearn}}

The \texttt{bnlearn} package offers many built-in choices for the required \texttt{test} input, including a likelihood ratio test (\texttt{mi-cg}) for mixed (Conditional Gaussian) variables similar to the one implemented in \texttt{micd}. A notable difference is that in \texttt{bnlearn}, it is assumed that continuous variables cannot be parents of discrete variables, and edges between continuous and discrete variables will always be oriented towards the continuous variable in the estimated CPDAG or MPDAG. If no test is specified in \texttt{pc.stable()} and the dataset contains both continuous and discrete variables, then \texttt{mi-cg} is automatically selected.

In the following code, we run \texttt{pc.stable()} on the same synthetic mixed data as before:

\begin{Shaded}
\begin{Highlighting}[]
\NormalTok{bnlearn\_fit\_mix }\OtherTok{\textless{}{-}} \FunctionTok{pc.stable}\NormalTok{(dat\_cross, }\AttributeTok{test =} \StringTok{"mi{-}cg"}\NormalTok{, }\AttributeTok{alpha =} \FloatTok{0.01}\NormalTok{)}
\FunctionTok{graphviz.plot}\NormalTok{(bnlearn\_fit\_mix, }\AttributeTok{shape =} \StringTok{"ellipse"}\NormalTok{)}
\end{Highlighting}
\end{Shaded}

\begin{figure}
\centering
\includegraphics{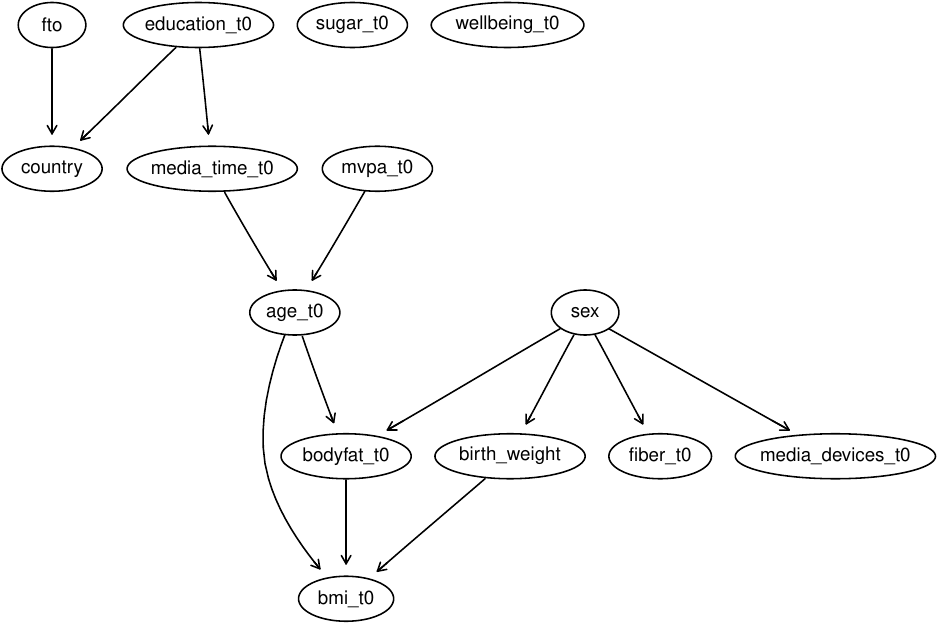}
\caption{\label{fig:bn-learn-mixed}Output from the \texttt{pc.stable()} function in \texttt{bnlearn}.}
\end{figure}

The output is shown in Figure \ref{fig:bn-learn-mixed}. Compared to Figure 6, one sees more directed edges (in fact, all edges are directed, but this does not have to be the case in general). This is because of how \texttt{bnlearn} handles conflicts -- it chooses one direction where \texttt{pcalg} would leave the edge undirected -- and because arrows between continuous and discrete variables are always oriented towards the continuous variable.

\hypertarget{tetrad-1}{%
\subsubsection{TETRAD}\label{tetrad-1}}

\begin{figure}
\centering
\includegraphics{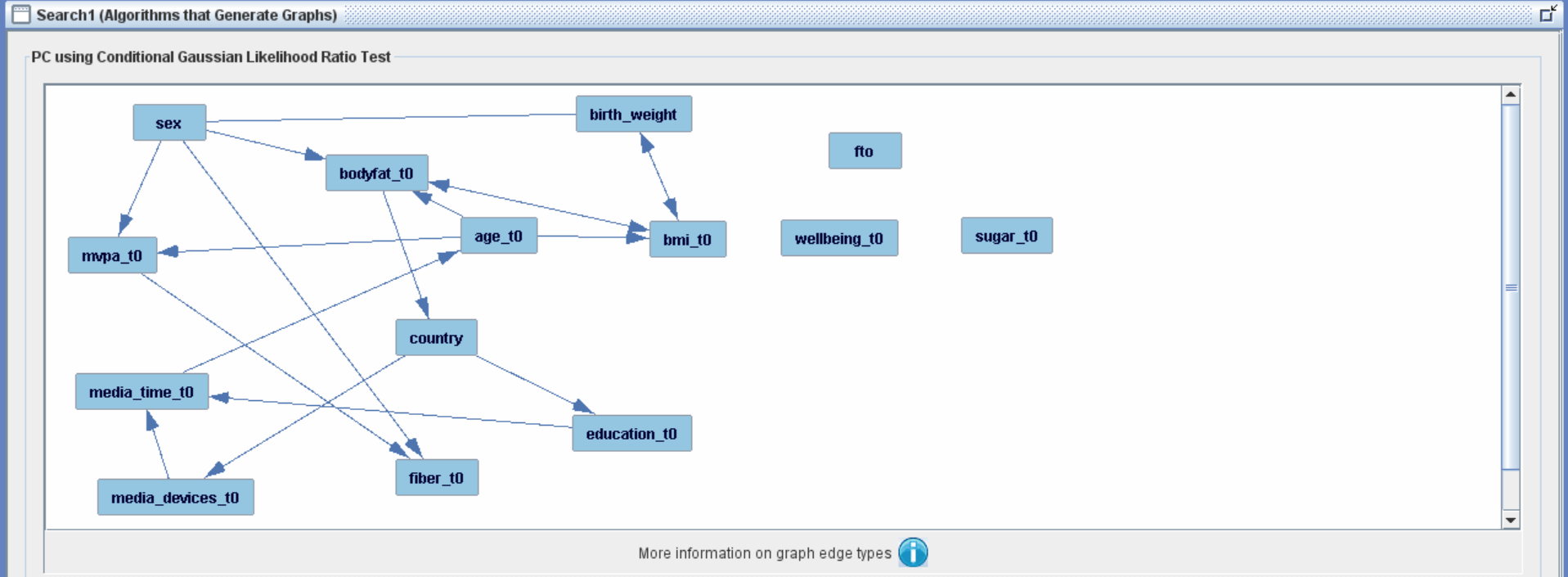}
\caption{\label{fig:tetrad-mixed}Output of the conditional Gaussian test with cross-sectional mixed data in TETRAD}
\end{figure}

Two tests for mixed data are currently implemented in TETRAD: the Conditional Gaussian test by \citet{AndrewsRamseyCooper2018} and the Degenerate Gaussian test by \citet{AndrewsRamseyCooper2019}, where categorical variables are internally converted into dummy variables and then treated as continuous. Figure \ref{fig:tetrad-search-algs} shows the ``Search'' box where the selection is made. The output of TETRAD using the Conditional Gaussian test and all the settings shown in Figure \ref{fig:tetrad-search-options} is in Figure \ref{fig:tetrad-mixed}.

Alternatively, data can be discretized or converted to a numeric scale by inserting a second ``Data'' box between the first ``Data'' box and the ``Search'' box, and choosing ``Discretize Dataset'' or ``Convert Numerical Discrete To Continuous'', respectively (screenshots not shown).

\hypertarget{time-ordering}{%
\subsection{Time ordering}\label{time-ordering}}

In longitudinal cohort studies, it is common that at least a subset of variables is measured at every study visit. The resulting partial temporal ordering of the variables should be accounted for in order to improve the result of causal discovery. There are different strategies for incorporating temporal constraints, depending on the assumptions one is willing to make about the data. In this guide, we focus on an option that is relatively close to the unmodified version of the PC algorithm: We restrict edges to be oriented in accordance with the flow of time, as described in greater detail in the next subsection. This strategy is followed, for example, in \citet{laBastideetal2014}.

Many other strategies for including time constraints have been discussed in the literature. For example, \citet{Rahmadietal2018} assume that the repeated measurements are equally spaced and that the causal relations among the repeatedly measured variables do not change over time, an assumption that is sometimes called stationarity. This allows them to pool the data from the different time points. A whole branch of the literature is concerned with causal discovery from time series data, where it is usually assumed that the data constitute instantaneous measurements, and hence directed edges within time points are not allowed (see \citet{Scutari2020} for a review). However, we believe that this assumption is not realistic for typical cohort data. Further, repeated measurements of the same variable should be considered as being time ordered; however, properly accounting for subject-level correlation is an active area of research.

\hypertarget{pcalg-and-add-ons-1}{%
\subsubsection{\texorpdfstring{\texttt{pcalg} and add-ons}{pcalg and add-ons}}\label{pcalg-and-add-ons-1}}

\begin{figure}
\centering
\includegraphics{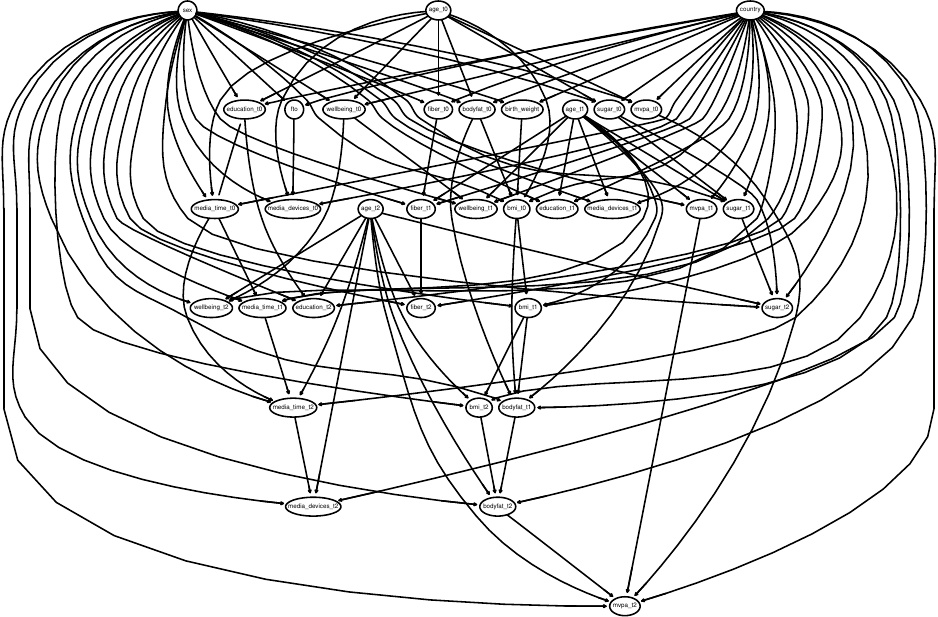}
\caption{\label{fig:pcalg-ordering}Output of \texttt{tpc()} after specifying the ordering of variables based on background knowledge.}
\end{figure}

Time-ordered measurements are accommodated by \texttt{tpc}. It implements the following two modifications first suggested by \citet{causationpredictionsearch}: Arrows between time points are always oriented from the earlier to the later time point, and conditional independence tests are generally restricted such that the variables in the conditioning set must not lie in the future of both variables whose conditional independence is to be tested. Further, it is possible in \texttt{tpc} to specify context variables. These are variables known to be exogenous, i.e., known to have no incoming edges. In our data example, we specify \texttt{context.all = c("country", "sex")} to mark these two variables as context variables and force edges from \texttt{country} and \texttt{sex} to all other variables into the graph. Further, we use \texttt{context.tier\ =\ c("age\_t0",\ "age\_t1",\ "age\_t2")} to mark these three variables as context variables that have edges into variables in their own tier, but not into variables from other tiers. The resulting graph estimate is shown in Figure \ref{fig:pcalg-ordering}.

\begin{Shaded}
\begin{Highlighting}[]
\DocumentationTok{\#\# specify tiers: (1) country, sex; (2) FTO gene; (3) birth weight;}
\DocumentationTok{\#\# (4) all t0 variables; (5) all t1 variables, (6) all t2 variables}
\NormalTok{tiers }\OtherTok{\textless{}{-}} \FunctionTok{rep}\NormalTok{( }\FunctionTok{c}\NormalTok{(}\DecValTok{1}\NormalTok{,}\DecValTok{2}\NormalTok{,}\DecValTok{3}\NormalTok{,}\DecValTok{4}\NormalTok{,}\DecValTok{5}\NormalTok{,}\DecValTok{6}\NormalTok{), }\AttributeTok{times =} \FunctionTok{c}\NormalTok{(}\DecValTok{2}\NormalTok{,}\DecValTok{1}\NormalTok{,}\DecValTok{1}\NormalTok{,}\DecValTok{10}\NormalTok{,}\DecValTok{10}\NormalTok{,}\DecValTok{10}\NormalTok{) )}

\NormalTok{pcalg\_fit }\OtherTok{\textless{}{-}} \FunctionTok{tpc}\NormalTok{(}\AttributeTok{suffStat =}\NormalTok{ dat\_cohort, }\AttributeTok{indepTest =}\NormalTok{ mixCItest, }\AttributeTok{alpha =} \FloatTok{0.01}\NormalTok{, }
                 \AttributeTok{labels =} \FunctionTok{colnames}\NormalTok{(dat\_cohort), }\AttributeTok{maj.rule =} \ConstantTok{TRUE}\NormalTok{,}
                 \AttributeTok{tiers =}\NormalTok{ tiers,}
                 \AttributeTok{context.all =} \FunctionTok{c}\NormalTok{(}\StringTok{"country"}\NormalTok{, }\StringTok{"sex"}\NormalTok{),}
                 \AttributeTok{context.tier =} \FunctionTok{c}\NormalTok{(}\StringTok{"age\_t0"}\NormalTok{, }\StringTok{"age\_t1"}\NormalTok{, }\StringTok{"age\_t2"}\NormalTok{))}
\FunctionTok{mygraph}\NormalTok{(pcalg\_fit)}
\end{Highlighting}
\end{Shaded}

\hypertarget{bnlearn-1}{%
\subsubsection{\texorpdfstring{\texttt{bnlearn}}{bnlearn}}\label{bnlearn-1}}

\begin{figure}
\centering
\includegraphics{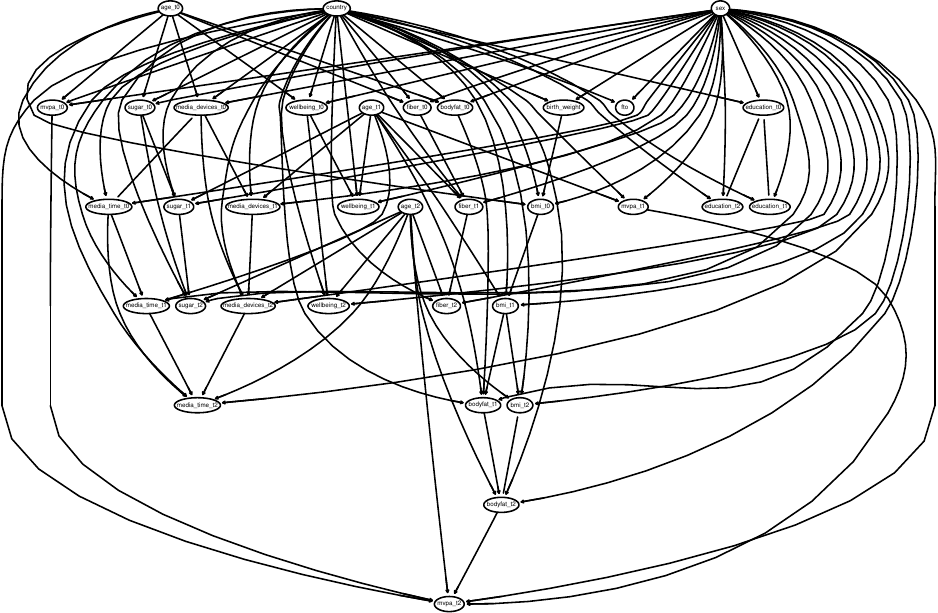}
\caption{\label{fig:bnlearn-ordering}Output of \texttt{pc.stable()} in \texttt{bnlearn} after specifying a time ordering of the variables based on black and white lists.}
\end{figure}

The \texttt{pc.stable()} function in \texttt{bnlearn} has a \texttt{blacklist} input, which is used to specify edges and edge directions that are forbidden. Below, we repeat the call to \texttt{pc.stable()} that we had in the prior ``mixed measurement scales'' section, but we use the full longitudinal data. We also specify that there cannot be any edges between variables that go backwards in time using the \texttt{tiers2blacklist()} function, that \texttt{country} and \texttt{sex} have edges into everything else, and that \texttt{age\_t0}, \texttt{age\_t1}, and \texttt{age\_t2} have edges into all (continuous) variables within their own tier, as in the previous section. Note that if we try to force the continuous age variables to have edges into the discrete education variables, we get an error. This is due to the assumption that continuous variables do not have discrete children, which \texttt{bnlearn} always makes when \texttt{test\ =\ ""mi-cg"} is specified. The output is shown in Figure \ref{fig:bnlearn-ordering}.

\begin{Shaded}
\begin{Highlighting}[]
\NormalTok{bl1 }\OtherTok{\textless{}{-}} \FunctionTok{tiers2blacklist}\NormalTok{(}\FunctionTok{split}\NormalTok{(}\FunctionTok{names}\NormalTok{(dat\_cohort), tiers))}
\NormalTok{bl2 }\OtherTok{\textless{}{-}} \FunctionTok{data.frame}\NormalTok{(}\AttributeTok{from =} \FunctionTok{names}\NormalTok{(dat\_cohort),}
                  \AttributeTok{to =} \FunctionTok{rep}\NormalTok{( }\FunctionTok{c}\NormalTok{(}\StringTok{"age\_t0"}\NormalTok{,}\StringTok{"age\_t1"}\NormalTok{,}\StringTok{"age\_t2"}\NormalTok{), }\AttributeTok{each =} \DecValTok{34}\NormalTok{ ))}
\NormalTok{bl3 }\OtherTok{\textless{}{-}} \FunctionTok{data.frame}\NormalTok{(}\AttributeTok{from =} \FunctionTok{c}\NormalTok{(}\StringTok{"country"}\NormalTok{,}\StringTok{"sex"}\NormalTok{), }\AttributeTok{to =} \FunctionTok{c}\NormalTok{(}\StringTok{"sex"}\NormalTok{,}\StringTok{"country"}\NormalTok{))}
\NormalTok{bl }\OtherTok{\textless{}{-}} \FunctionTok{rbind}\NormalTok{(bl1, bl2, bl3)}
\NormalTok{wl1 }\OtherTok{\textless{}{-}} \FunctionTok{data.frame}\NormalTok{(}\AttributeTok{from =} \FunctionTok{rep}\NormalTok{( }\FunctionTok{c}\NormalTok{(}\StringTok{"country"}\NormalTok{,}\StringTok{"sex"}\NormalTok{), }\AttributeTok{each =} \DecValTok{29}\NormalTok{ ),}
                  \AttributeTok{to =} \FunctionTok{names}\NormalTok{(dat\_cohort)[}\SpecialCharTok{{-}}\FunctionTok{c}\NormalTok{(}\DecValTok{1}\NormalTok{,}\DecValTok{2}\NormalTok{,}\DecValTok{5}\NormalTok{,}\DecValTok{15}\NormalTok{,}\DecValTok{25}\NormalTok{)])}
\NormalTok{wl2 }\OtherTok{\textless{}{-}} \FunctionTok{data.frame}\NormalTok{(}\AttributeTok{from =} \StringTok{"age\_t0"}\NormalTok{,}
                  \AttributeTok{to =} \FunctionTok{c}\NormalTok{(}\StringTok{"bmi\_t0"}\NormalTok{,}\StringTok{"bodyfat\_t0"}\NormalTok{,}\StringTok{"fiber\_t0"}\NormalTok{, }\StringTok{"media\_devices\_t0"}\NormalTok{,}
                       \StringTok{"media\_time\_t0"}\NormalTok{,}\StringTok{"mvpa\_t0"}\NormalTok{,}\StringTok{"sugar\_t0"}\NormalTok{,}\StringTok{"wellbeing\_t0"}\NormalTok{))}
\NormalTok{wl3 }\OtherTok{\textless{}{-}} \FunctionTok{data.frame}\NormalTok{(}\AttributeTok{from =} \StringTok{"age\_t1"}\NormalTok{,}
                  \AttributeTok{to =} \FunctionTok{c}\NormalTok{(}\StringTok{"bmi\_t1"}\NormalTok{,}\StringTok{"bodyfat\_t1"}\NormalTok{,}\StringTok{"fiber\_t1"}\NormalTok{,}\StringTok{"media\_devices\_t1"}\NormalTok{,}
                       \StringTok{"media\_time\_t1"}\NormalTok{,}\StringTok{"mvpa\_t1"}\NormalTok{,}\StringTok{"sugar\_t1"}\NormalTok{,}\StringTok{"wellbeing\_t1"}\NormalTok{))}
\NormalTok{wl4 }\OtherTok{\textless{}{-}} \FunctionTok{data.frame}\NormalTok{(}\AttributeTok{from =} \StringTok{"age\_t2"}\NormalTok{,}
                  \AttributeTok{to =} \FunctionTok{c}\NormalTok{(}\StringTok{"bmi\_t2"}\NormalTok{,}\StringTok{"bodyfat\_t2"}\NormalTok{,}\StringTok{"fiber\_t2"}\NormalTok{,}\StringTok{"media\_devices\_t2"}\NormalTok{,}
                       \StringTok{"media\_time\_t2"}\NormalTok{,}\StringTok{"mvpa\_t2"}\NormalTok{,}\StringTok{"sugar\_t2"}\NormalTok{,}\StringTok{"wellbeing\_t2"}\NormalTok{))}
\NormalTok{wl }\OtherTok{\textless{}{-}} \FunctionTok{rbind}\NormalTok{(wl1, wl2, wl3, wl4)}

\CommentTok{\# causal discovery}
\NormalTok{bnlearn\_fit }\OtherTok{\textless{}{-}} \FunctionTok{pc.stable}\NormalTok{(dat\_cohort, }\AttributeTok{alpha =} \FloatTok{0.01}\NormalTok{,}
                         \AttributeTok{blacklist =}\NormalTok{ bl, }\AttributeTok{whitelist =}\NormalTok{ wl)}
\FunctionTok{graphviz.plot}\NormalTok{(bnlearn\_fit, }\AttributeTok{shape =} \StringTok{"ellipse"}\NormalTok{)}
\end{Highlighting}
\end{Shaded}

\hypertarget{tetrad-2}{%
\subsubsection{TETRAD}\label{tetrad-2}}

\begin{figure}
\centering
\includegraphics{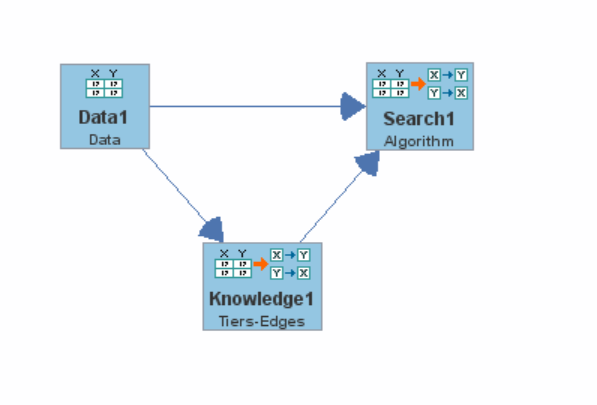}
\caption{\label{fig:tetrad-knowledge}Adding a ``Knowledge'' box in TETRAD.}
\end{figure}

In TETRAD, users can ensure correct temporal ordering by adding a ``Knowledge'' box to the causal search pipeline, see Figure \ref{fig:tetrad-knowledge}. This is equivalent to specifying the tiers in \texttt{tpc}.
``Knowledge'' boxes also allow users to specify required and forbidden edges between variables, much like the white and black lists of \texttt{bnlearn}.

\begin{figure}
\centering
\includegraphics{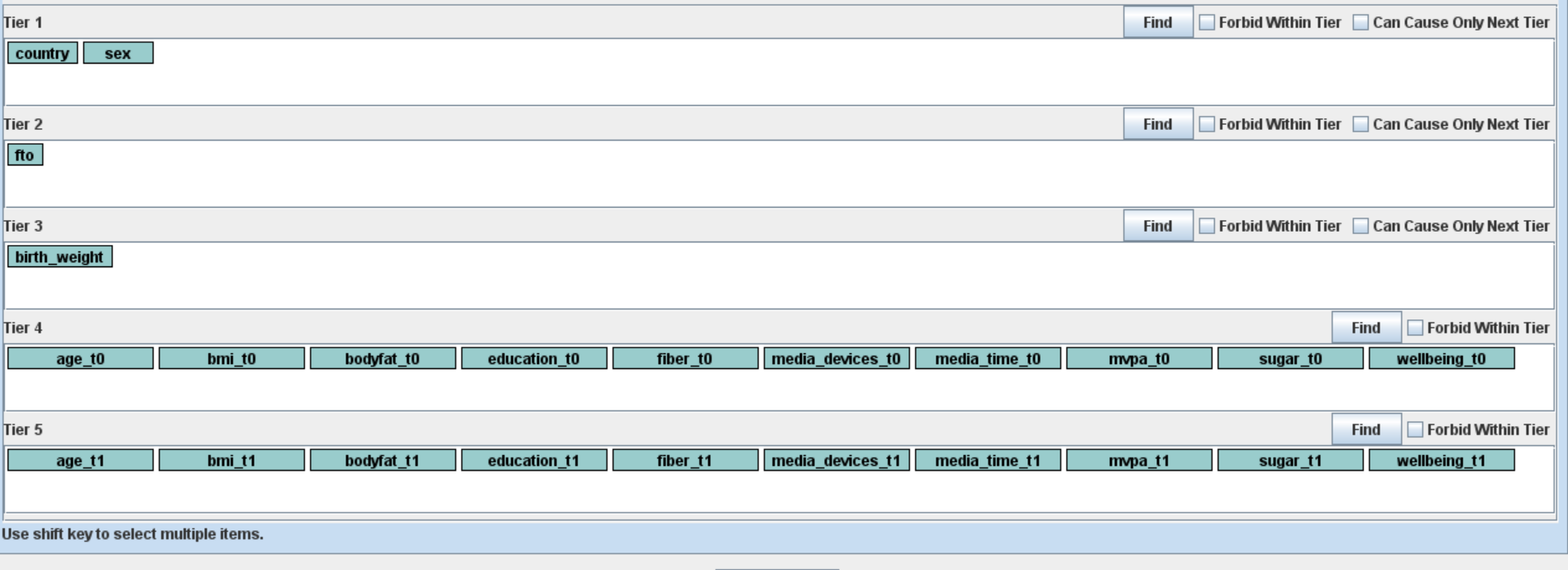}
\caption{\label{fig:tetrad-tiers}TETRAD allows users to specify tiers, just like \texttt{bnlearn} and \texttt{tpc}.}
\end{figure}

Even though our data has three time points, we will illustrate the steps needed in TETRAD using only the first two time points. Sorting variables into tiers is a straightforward process -- one need only click on a variable in the ``Not in Tier'' box, and drag it to the appropriate Tier, see Figure \ref{fig:tetrad-tiers}. If desired, there are additional options for each tier, like specifying that there are no causal relationships between variables within a tier (``Forbid Within Tier'') or that variables within a particular tier can only have causal relationships with variables in the next tier (``Can Cause Only Next Tier'').

\begin{figure}
\centering
\includegraphics{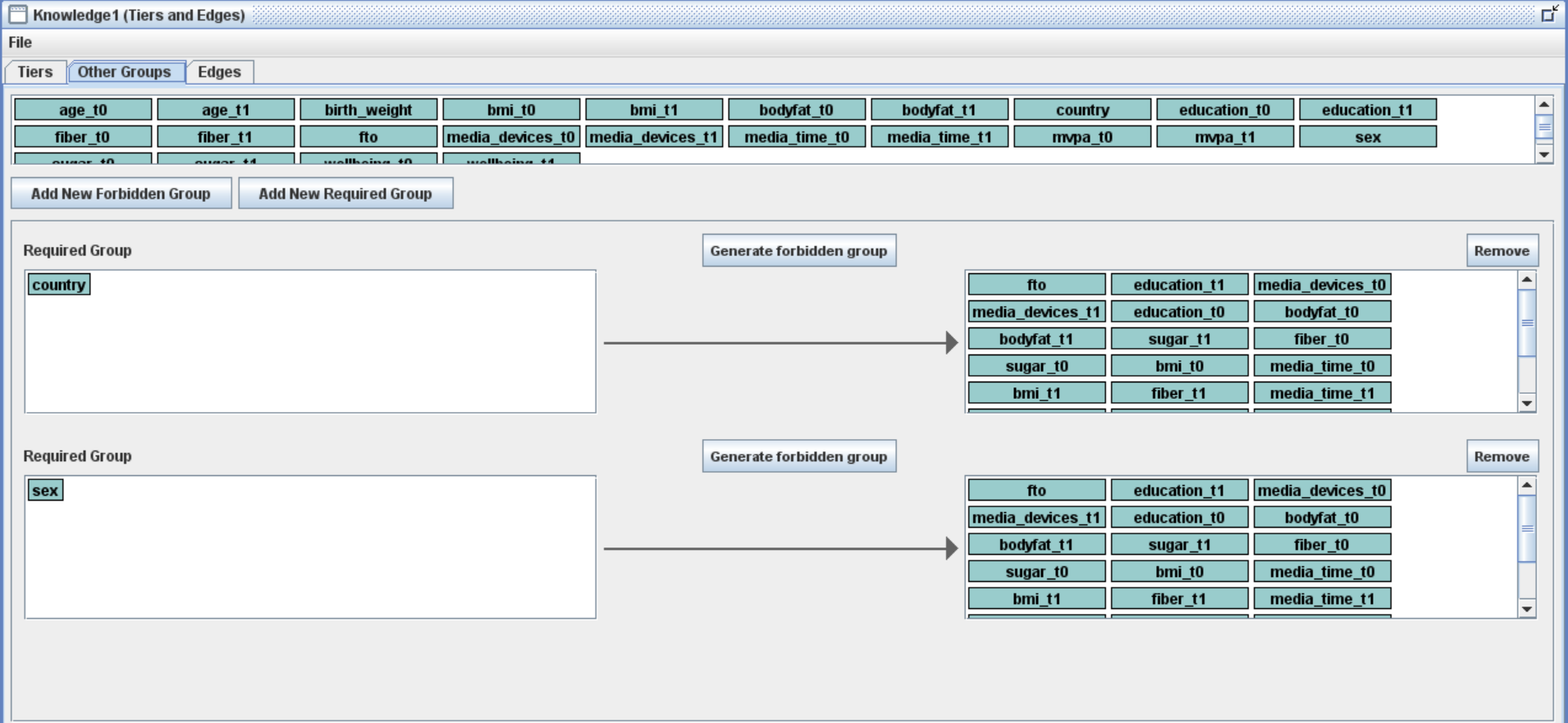}
\caption{\label{fig:tetrad-required}TETRAD also allows for forbidden and required relationships to be specified. This is similar to the \texttt{context.all} and \texttt{context.tier} arguments of \texttt{tpc()} and the \texttt{blacklist} and \texttt{whitelist} arguments of \texttt{pc.stable()}.}
\end{figure}

In our data, we believe that country and sex influence every other variable. To tell TETRAD this information, we use a ``required'' group, see Figure \ref{fig:tetrad-required}. Just like in the prior analyses with \texttt{tpc} and \texttt{bnlearn}, we also specify that the variables \texttt{age\_t0} and \texttt{age\_t1} have edges into all other variables in their respective tiers, but not outside of them, and that country, sex, and the age variables do not influence each other (screenshots not shown).

It turns out that without these knowledge-based specifications, TETRAD will not produce any output as it struggles to perform all of the required conditional independence tests. With two time points and 24 variables, we found that TETRAD took approximately 24-36 hours to complete the search. Users with many variables and/or time points could experience substantial run times, and this should be factored into any analysis plans.

\begin{figure}
\centering
\includegraphics{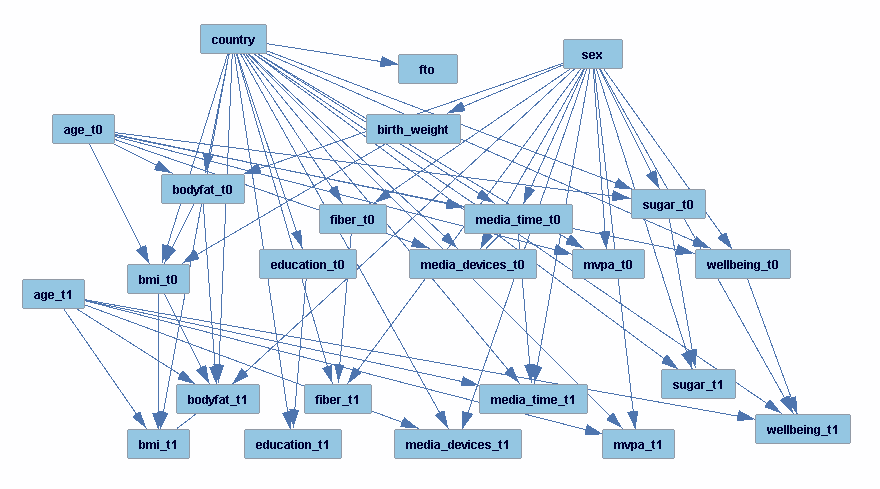}
\caption{\label{fig:tetrad-timeordered-output}TETRAD output after specifying a time ordering between variables using tiers, required groups, and forbidden groups.}
\end{figure}

Figure \ref{fig:tetrad-timeordered-output} is the output of our search. Overall, TETRAD returned many edges between variables; however, we also note that not all of the required edges are found in the graph (e.g., there is no edge between \texttt{age\_t0} and \texttt{fiber\_t0} even though we required it). This is a known issue with TETRAD, and the developers are actively addressing it. In the meantime, we advise users of TETRAD to review all output carefully to ensure it matches what was expected given one's background knowledge.

\hypertarget{missing-data}{%
\subsection{Missing data}\label{missing-data}}

Whenever a participant in a cohort study misses a study visit, drops out of the study, or cannot complete all scheduled study visit assessments, missing data points are created. Until recently, most causal discovery algorithms have ignored missing values and only ad-hoc methods were applied. A simple but problematic method is to eliminate missing values in a pre-processing step, by either removing any data rows with missing values (`list-wise deletion' or `complete case analysis') or by replacing missing values e.g.,~by the column mode or median (`single imputation'). Both are problematic: List-wise deletion induces a selection bias in general under most missingness mechanisms. The problem with single imputation is that the uncertainty in the imputed values is not accounted for, which leads to biased results of the conditional independence tests in PC and other algorithms.

An alternative to list-wise deletion is test-wise deletion, where the decision to remove or retain an observation is made for each conditional independence test separately \citep[\citet{Tuetal2020}, \citet{Witteetal2021}]{StroblVisweswaranSpirtes2018}. For example, if the data row for individual \(i\) is complete except for missing well-being, then this data row is removed for all conditional independence tests that include the well-being variable, but is retained otherwise. In test-wise deletion, more of the overall sample can be retained, which can lead to a higher power for some tests compared to list-wise deletion. Just as list-wise deletion, however, test-wise deletion suffers from a selection bias for some missingness mechanisms \citep[\citet{Witteetal2021}]{Tuetal2020}.

A valid imputation procedure under the missing at random (MAR) assumption \citep{Rubin1976} is multiple imputation, where each missing value is replaced by several different plausible values to account for the uncertainty about the missing values \citep{Rubin1987}. The conditional independence tests in the causal discovery algorithm then need to be adapted in order to handle the multiply imputed data. This is described in more detail in \citet{Foraitaetal2020} and \citet{Witteetal2021}. Multiple imputation relies on the MAR assumption and on correctly specified imputation models.

Based on the missing data patterns seen in IDEFICS/I.Family, we created a dataset with 1-10\% missing data for most variables. Exceptions were that country, age, sex, and body mass index were never missing, while fiber intake and physical activity measurements were missing 50-60\% of the time. In order to reduce the runtime of the multiple imputation procedure, we only consider baseline and time 0 in the data examples below.

\hypertarget{pcalg-and-add-ons-2}{%
\subsubsection{\texorpdfstring{\texttt{pcalg} and add-ons}{pcalg and add-ons}}\label{pcalg-and-add-ons-2}}

\begin{figure}
\centering
\includegraphics{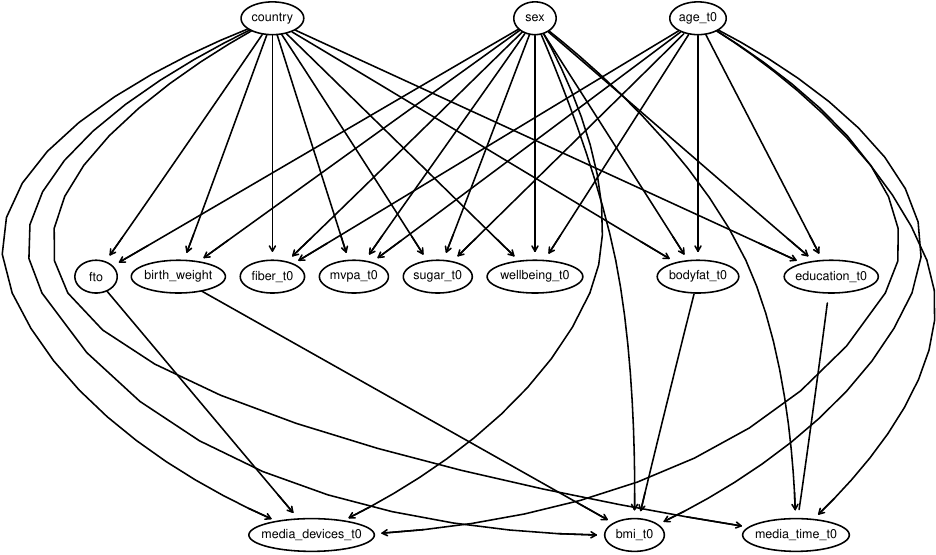}
\caption{\label{fig:pcalg-missing-twd}Output of \texttt{tpc()} when the input data has missing values. The function applies test-wise deletion, specified via \texttt{indepTest\ =\ mixCItwd}.}
\end{figure}

\begin{figure}
\centering
\includegraphics{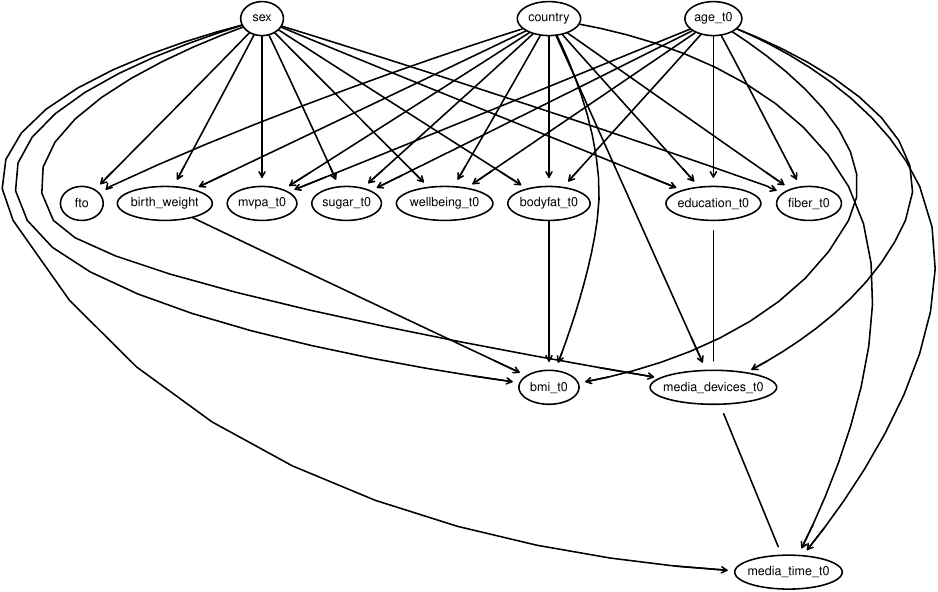}
\caption{\label{fig:pcalg-missing-MI}Output of \texttt{tpc()} when the input is a list of multiply imputed datasets generated via random forest imputation.}
\end{figure}

The \texttt{micd} add-on to \texttt{pcalg} offers functions for performing test-wise deletion and multiple imputation. Test-wise deletion is performed by simply replacing \texttt{gaussCItest}, \texttt{disCItest}, or \texttt{mixCItest} by \texttt{gaussCItwd}, \texttt{disCItwd}, or \texttt{mixCItwd}, respectively. For a multiple imputation analysis, the first step is to generate multiply imputed data using e.g.,~the \texttt{mice} package \citep{R_mice} with the random forest option for generating the imputations (\texttt{method = "rf"}; this may take several minutes to run). In the second step, \texttt{pc()} is run with \texttt{indepTest\ =\ gaussMItest}, \texttt{disMItest}, or \texttt{mixMItest}, respectively. Figures \ref{fig:pcalg-missing-twd} and \ref{fig:pcalg-missing-MI} show the graphs estimated using \texttt{mixCItwd} or \texttt{mixMItest}, respectively, produced by the following code:

\begin{Shaded}
\begin{Highlighting}[]
\DocumentationTok{\#\# load incomplete data and create cross{-}sectional version of the dataset}
\FunctionTok{data}\NormalTok{(}\StringTok{"dat\_cohort\_mis"}\NormalTok{)}
\NormalTok{dat\_miss\_cross }\OtherTok{\textless{}{-}}\NormalTok{ dat\_cohort\_mis[ ,}\DecValTok{1}\SpecialCharTok{:}\DecValTok{14}\NormalTok{]}

\NormalTok{pcalg\_fit\_twd }\OtherTok{\textless{}{-}} \FunctionTok{tpc}\NormalTok{(}\AttributeTok{suffStat =}\NormalTok{ dat\_miss\_cross, }\AttributeTok{indepTest =}\NormalTok{ mixCItwd,}
                     \AttributeTok{alpha =} \FloatTok{0.01}\NormalTok{, }\AttributeTok{labels =} \FunctionTok{colnames}\NormalTok{(dat\_miss\_cross),}
                     \AttributeTok{maj.rule =} \ConstantTok{TRUE}\NormalTok{,}
                     \AttributeTok{tiers =}\NormalTok{ tiers[}\DecValTok{1}\SpecialCharTok{:}\DecValTok{14}\NormalTok{],}
                     \AttributeTok{context.all =} \FunctionTok{c}\NormalTok{(}\StringTok{"country"}\NormalTok{, }\StringTok{"sex"}\NormalTok{),}
                     \AttributeTok{context.tier =} \FunctionTok{c}\NormalTok{(}\StringTok{"age\_t0"}\NormalTok{))}

\FunctionTok{mygraph}\NormalTok{(pcalg\_fit\_twd)}
\end{Highlighting}
\end{Shaded}

\begin{Shaded}
\begin{Highlighting}[]
\DocumentationTok{\#\# install required package}
\CommentTok{\# install.packages("mice")}

\DocumentationTok{\#\# load required package}
\FunctionTok{library}\NormalTok{(mice)}

\DocumentationTok{\#\# generate multiply imputed data using random forest imputation}
\NormalTok{mi\_object }\OtherTok{\textless{}{-}} \FunctionTok{mice}\NormalTok{(dat\_miss\_cross, }\AttributeTok{m =} \DecValTok{10}\NormalTok{, }\AttributeTok{method =} \StringTok{"rf"}\NormalTok{, }\AttributeTok{print =} \ConstantTok{FALSE}\NormalTok{)}
\NormalTok{mi\_dat }\OtherTok{\textless{}{-}} \FunctionTok{complete}\NormalTok{(mi\_object, }\AttributeTok{action =} \StringTok{"all"}\NormalTok{)}

\DocumentationTok{\#\# apply PC}
\NormalTok{pcalg\_fit\_mi }\OtherTok{\textless{}{-}} \FunctionTok{tpc}\NormalTok{(}\AttributeTok{suffStat =}\NormalTok{ mi\_dat, }\AttributeTok{indepTest =}\NormalTok{ mixMItest,}
                    \AttributeTok{alpha =} \FloatTok{0.01}\NormalTok{, }\AttributeTok{labels =} \FunctionTok{colnames}\NormalTok{(dat\_miss\_cross),}
                    \AttributeTok{maj.rule =} \ConstantTok{TRUE}\NormalTok{,}
                    \AttributeTok{tiers =}\NormalTok{ tiers[}\DecValTok{1}\SpecialCharTok{:}\DecValTok{14}\NormalTok{],}
                    \AttributeTok{context.all =} \FunctionTok{c}\NormalTok{(}\StringTok{"country"}\NormalTok{, }\StringTok{"sex"}\NormalTok{),}
                    \AttributeTok{context.tier =} \FunctionTok{c}\NormalTok{(}\StringTok{"age\_t0"}\NormalTok{))}

\FunctionTok{mygraph}\NormalTok{(pcalg\_fit\_mi)}
\end{Highlighting}
\end{Shaded}

\hypertarget{bnlearn-2}{%
\subsubsection{\texorpdfstring{\texttt{bnlearn}}{bnlearn}}\label{bnlearn-2}}

\begin{figure}
\centering
\includegraphics{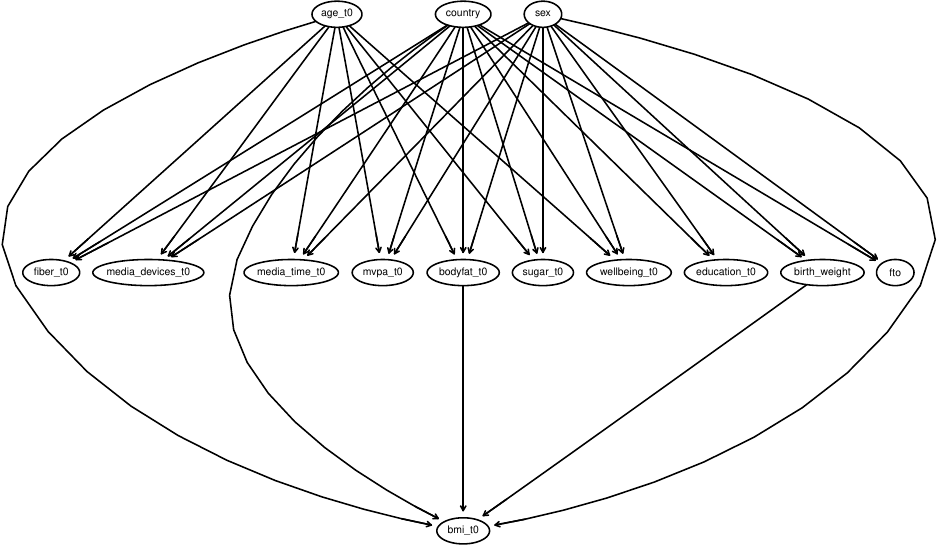}
\caption{\label{fig:bnlearn-missing}Output of \texttt{pc.stable()} in \texttt{bnlearn} when the input data has missing values. Here, the function performs test-wise deletion.}
\end{figure}

In \texttt{bnlearn}, the \texttt{pc.stable()} function performs test-wise deletion if the data is incomplete. Here, we repeat our call to \texttt{pc.stable()} from before, but use our dataset with missing data. The resulting graph estimate is shown in Figure \ref{fig:bnlearn-missing}. It is different from the graph estimated using test-wise deletion in \texttt{tpc}, due to the different implementations of the Conditional Gaussian conditional independence test and the different ways in which the packages handle conflicts.

\begin{Shaded}
\begin{Highlighting}[]
\NormalTok{bl1 }\OtherTok{\textless{}{-}} \FunctionTok{tiers2blacklist}\NormalTok{(}\FunctionTok{split}\NormalTok{(}\FunctionTok{names}\NormalTok{(dat\_miss\_cross), tiers[}\DecValTok{1}\SpecialCharTok{:}\DecValTok{14}\NormalTok{]))}
\NormalTok{bl2 }\OtherTok{\textless{}{-}} \FunctionTok{data.frame}\NormalTok{(}\AttributeTok{from =} \FunctionTok{names}\NormalTok{(dat\_miss\_cross), }\AttributeTok{to =} \StringTok{"age\_t0"}\NormalTok{)}
\NormalTok{bl3 }\OtherTok{\textless{}{-}} \FunctionTok{data.frame}\NormalTok{(}\AttributeTok{from =} \FunctionTok{c}\NormalTok{(}\StringTok{"country"}\NormalTok{,}\StringTok{"sex"}\NormalTok{), }\AttributeTok{to =} \FunctionTok{c}\NormalTok{(}\StringTok{"sex"}\NormalTok{,}\StringTok{"country"}\NormalTok{))}
\NormalTok{bl }\OtherTok{\textless{}{-}} \FunctionTok{rbind}\NormalTok{(bl1, bl2, bl3)}
\NormalTok{wl1 }\OtherTok{\textless{}{-}} \FunctionTok{data.frame}\NormalTok{(}\AttributeTok{from =} \FunctionTok{rep}\NormalTok{( }\FunctionTok{c}\NormalTok{(}\StringTok{"country"}\NormalTok{,}\StringTok{"sex"}\NormalTok{), }\AttributeTok{each =} \DecValTok{11}\NormalTok{ ),}
                  \AttributeTok{to =} \FunctionTok{names}\NormalTok{(dat\_miss\_cross)[}\SpecialCharTok{{-}}\FunctionTok{c}\NormalTok{(}\DecValTok{1}\NormalTok{,}\DecValTok{2}\NormalTok{,}\DecValTok{5}\NormalTok{)])}
\NormalTok{wl2 }\OtherTok{\textless{}{-}} \FunctionTok{data.frame}\NormalTok{(}\AttributeTok{from =} \StringTok{"age\_t0"}\NormalTok{,}
                  \AttributeTok{to =} \FunctionTok{c}\NormalTok{(}\StringTok{"bmi\_t0"}\NormalTok{,}\StringTok{"bodyfat\_t0"}\NormalTok{,}\StringTok{"fiber\_t0"}\NormalTok{, }\StringTok{"media\_devices\_t0"}\NormalTok{,}
                       \StringTok{"media\_time\_t0"}\NormalTok{,}\StringTok{"mvpa\_t0"}\NormalTok{,}\StringTok{"sugar\_t0"}\NormalTok{,}\StringTok{"wellbeing\_t0"}\NormalTok{))}
\NormalTok{wl }\OtherTok{\textless{}{-}} \FunctionTok{rbind}\NormalTok{(wl1, wl2)}

\NormalTok{bnlearn\_fit\_twd }\OtherTok{\textless{}{-}} \FunctionTok{pc.stable}\NormalTok{(dat\_miss\_cross, }\AttributeTok{alpha =} \FloatTok{0.01}\NormalTok{,}
                             \AttributeTok{blacklist =}\NormalTok{ bl, }\AttributeTok{whitelist =}\NormalTok{ wl)}
\FunctionTok{graphviz.plot}\NormalTok{(bnlearn\_fit\_twd, }\AttributeTok{shape =} \StringTok{"ellipse"}\NormalTok{)}
\end{Highlighting}
\end{Shaded}

For score-based causal discovery, \texttt{bnlearn} offers the \texttt{structural.EM()} function, which implements the Structural Expectation Maximization algorithm. It deals with missing values by iteratively completing the data and searching for the graphical structure \citep{Friedman1997}.

\hypertarget{tetrad-3}{%
\subsubsection{TETRAD}\label{tetrad-3}}

\begin{figure}
\centering
\includegraphics{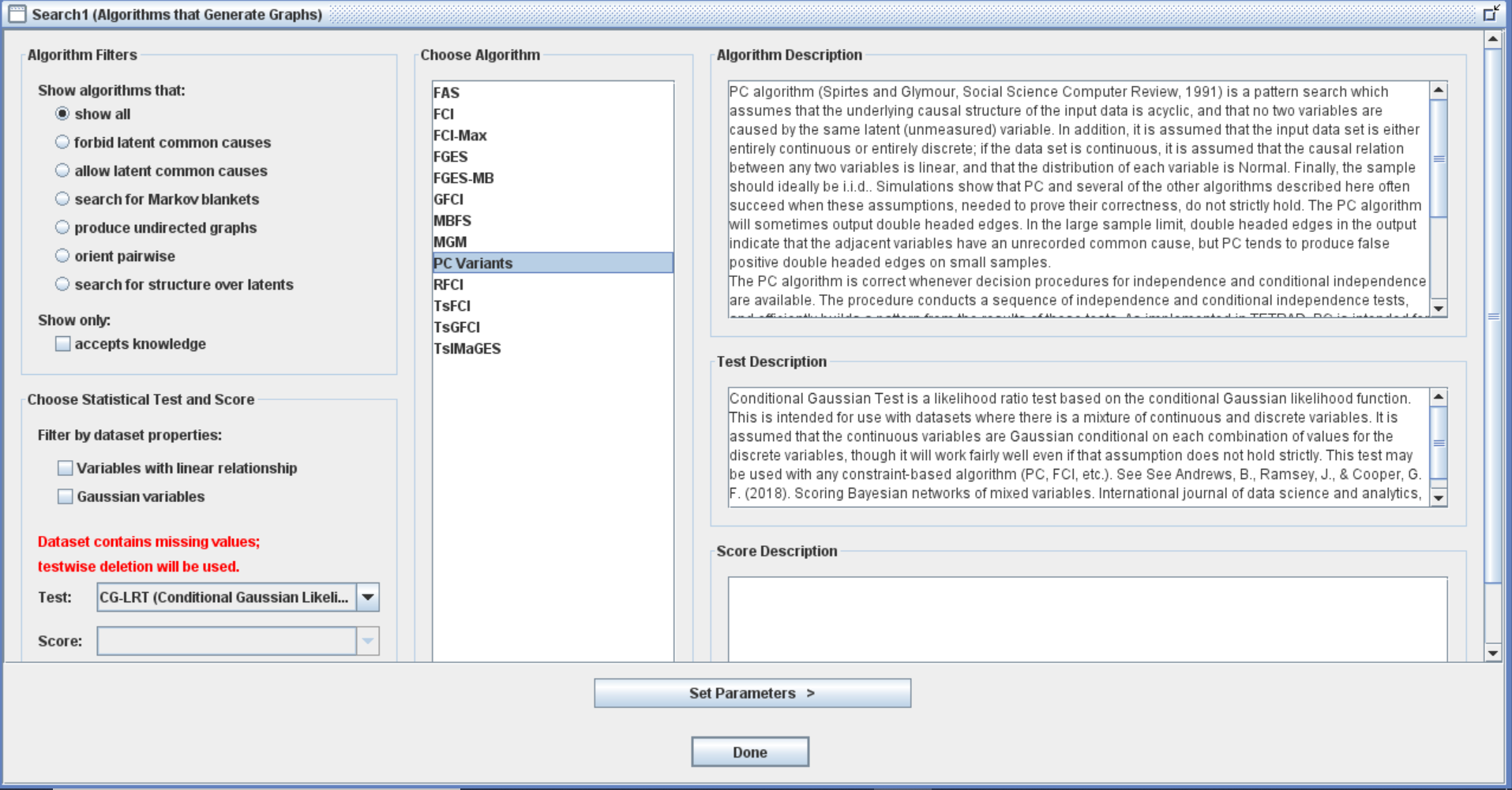}
\caption{\label{fig:search-algorithm-missing-warning}TETRAD warns users that test-wise deletion will be used when the data have missing values}
\end{figure}

\begin{figure}
\centering
\includegraphics{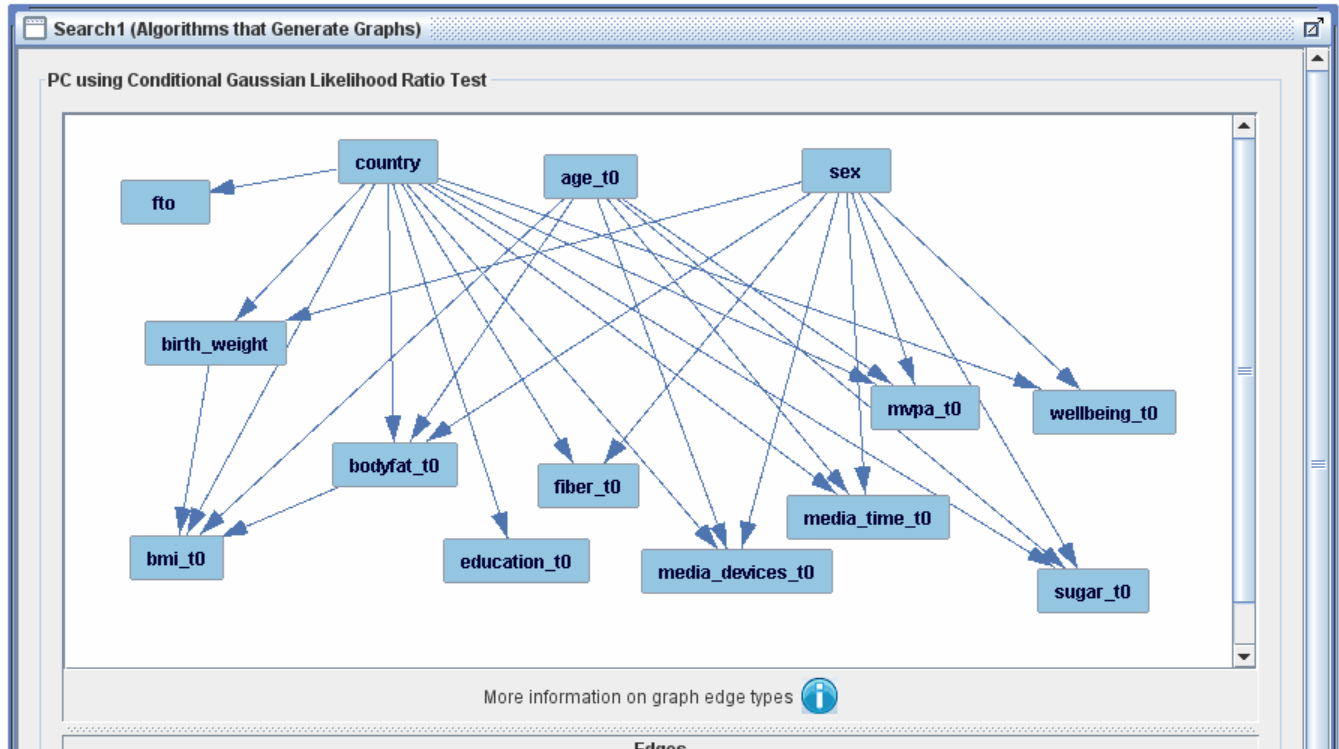}
\caption{\label{fig:tetrad-missing}Output of TETRAD when the input data has missing values. TETRAD performs test-wise deletion.}
\end{figure}

As \texttt{bnlearn}, TETRAD performs test-wise deletion by default if the data is incomplete, and it warns users of this when selecting a search algorithm (see Figure \ref{fig:search-algorithm-missing-warning}). Figure \ref{fig:tetrad-missing} shows the graph estimated by TETRAD using the same incomplete data as before. Alternatively, list-wise deletion or single imputation may be performed by inserting a second ``Data'' box between the ``Data'' box and the ``Search'' box, and choosing one of the options listed under `Missing Values' (not recommended). Again, one will notice that not all required edges are present in the output graph, and we reiterate the importance of checking all TETRAD output against background knowledge to ensure that the output theoretically makes sense.

\hypertarget{additional-topic-causal-effect-estimation}{%
\subsection{Additional topic: Causal effect estimation}\label{additional-topic-causal-effect-estimation}}

So far in this guide, we have focused on how software can be used to learn the structure of a graph in the face of common issues with observational data. In many studies, however, one would ideally like to supplement this by quantifying (i.e., estimating) the overall (direct or indirect) causal effects that correspond to directed edges and directed paths found. To that end, causal discovery algorithms can be extended by the ``Intervention calculus when the DAG is Absent'' (IDA) algorithm to estimate causal effects. The original IDA algorithm assumes that one's data was generated from a multivariate Gaussian distribution, and also assumes that the data is faithful to an underlying causal DAG. Because the equivalence class represented by a CPDAG or MPDAG contains more than one DAG, the IDA algorithm does not estimate a single causal effect of \(X\) on \(Y\); rather, it estimates multiple possible causal effects. These can be summarized as appropriate for the study question of interest (e.g., taking the minimum and maximum provides bounds for the causal effect).

In \texttt{pcalg}, the \texttt{ida()} function performs the IDA algorithm. The function requires that a valid CPDAG or MPDAG be provided as input. If the graph is not valid, one will receive an error and must manually alter the graph based on subject matter knowledge until it is valid.

\hypertarget{discussion}{%
\section{Discussion}\label{discussion}}

Causal discovery is a useful tool that is not as widely known and implemented in public health and related fields, possibly because applying existing causal discovery tools to observational cohort data can be not straightforward. In this guide, we have provided details on how to use three software packages -- \texttt{pcalg}, \texttt{bnlearn}, and TETRAD -- to learn the causal structure between variables in the presence of mixed data, time ordering, and missing data. While these packages can be used in these situations, their default capabilities are limited. Therefore, we also demonstrated how the new packages \texttt{tpc} and \texttt{micd} provide more functionality for causal discovery using cohort data, particularly when it comes to time ordered variables and missing data. Where possible, we highlighted key differences and similarities between the packages.

As mentioned briefly at the beginning of this guide, Python wrappers and packages exist for causal discovery as well. In fact, there are Python implementations of the \texttt{pcalg} and \texttt{bnlearn} packages that are equivalent to the implementations in \texttt{R}. We have not tested these Python implementations and have therefore not included them in this guide; however, we encourage users more comfortable in Python programming to learn more about them.

In terms of practical recommendations, we encourage analysts to choose software that meets their needs. For example, the \texttt{bnlearn} package has straightforward black and white list options that can ensure that time ordering between variables is reflected in any output graph. At the same time, \texttt{bnlearn} has limited functionality for resolving edge conflicts, which may lead analysts to prefer another software. Similarly, the \texttt{tpc} package relies on much of the functionality of \texttt{pcalg}, while also providing additional functions. If the analyst prefers or requires a graphical interface for causal discovery, the only available software is TETRAD. Regardless of which software is chosen, we also encourage analysts to consider performing their causal discovery analysis with multiple packages. Obtaining similar results across packages could lend plausibility to the overall results, while any differences seen could help pinpoint potential issues or help with the interpretation of the findings.

\renewcommand\refname{References}
  \bibliography{DFGbibtex.bib}
\end{document}